\documentclass[twocolumn,english,conference]{IEEEtran}
\usepackage[T1]{fontenc}
\usepackage{xcolor}
\usepackage{babel}
\usepackage{array}
\usepackage{verbatim}
\usepackage{float}
\usepackage{textcomp}
\usepackage{multirow}
\usepackage{amsmath}
\usepackage{amssymb}
\usepackage{stackrel}
\usepackage{graphicx}
\usepackage{wasysym}
\usepackage[unicode=true,
 bookmarks=true,bookmarksnumbered=true,bookmarksopen=true,bookmarksopenlevel=1,
 breaklinks=false,pdfborder={0 0 0},pdfborderstyle={},backref=false,colorlinks=false]
 {hyperref}
\hypersetup{pdftitle={Your Title},
 pdfauthor={Your Name},
 pdfpagelayout=OneColumn, pdfnewwindow=true, pdfstartview=XYZ, plainpages=false}

\makeatletter

\newcommand{\lyxmathsym}[1]{\ifmmode\begingroup\def\b@ld{bold}
  \text{\ifx\math@version\b@ld\bfseries\fi#1}\endgroup\else#1\fi}

\providecommand{\tabularnewline}{\\}
\newcommand{\lyxdot}{.}

\floatstyle{ruled}
\newfloat{algorithm}{tbp}{loa}
\providecommand{\algorithmname}{Algorithm}
\floatname{algorithm}{\protect\algorithmname}

\IEEEoverridecommandlockouts
\ifCLASSOPTIONcompsoc
\usepackage[caption=false,font=normalsize,labelfont=sf,textfont=sf]{subfig}
\else
\usepackage[caption=false,font=footnotesize]{subfig}
\fi
\usepackage{cite}
\usepackage{amsthm}

\newtheorem{proposition}{\textbf{Proposition}}
\usepackage{algorithm}
\usepackage{amsmath}

\usepackage{cite}
\usepackage{bm}
\usepackage{algorithmic}
\usepackage{algorithm}
\usepackage{graphicx}
\renewcommand{\fnum@figure}{Fig.~\thefigure}
\IEEEoverridecommandlockouts

\makeatother

\begin{document}
\title{3C Resources Joint Allocation for Time-Deterministic Remote Sensing
Image Backhaul in the Space-Ground Integrated Network}
\author{\IEEEauthorblockN{Chongxiao Cai, Yan Zhu,\emph{ Member, IEEE}, Min
Sheng, \emph{Fellow, IEEE}, Jiandong Li, \emph{Fellow, IEEE}, \\Yan
Shi, \emph{Member, IEEE}, Di Zhou, \emph{Member, IEEE}, Ziwen Xie,
and Chen Zhang }\IEEEauthorblockA{\textsuperscript{}State Key
Laboratory of Integrated Service Networks Xidian University, Xi'an,
Shaanxi, 710071, China\\
Email: chongxiaocai@stu.xidian.edu.cn;}}
\maketitle
\begin{abstract}
Low-Earth-orbit (LEO) satellites assist observation satellites (OSs)
to compress and backhaul more time-determined images (TDI) has become
a new paradigm, which is used to enhance the timeout caused by the
limited computing resources of OSs. However, how to capture the time-varying
and dynamic characteristics of multi-dimensional resources is challenging
for efficient collaborative scheduling. Motivated by this factor,
we design a highly succinct multi-dimensional resource time-expanded
graph (MDR-TEG) modell. Specifically, by employing a slots division
mechanism and introducing an external virtual node, the time-varying
communication, caching, and computing (3C) resources are depicted
in low complexity by the link weights within, between, and outside
the slots. Based on the MDR-TEG, the maximizing successful transmission
ratio of TDI (MSTR-TDI) is modeled as a mixed integer linear programming
(MILP) problem. Which further relaxed decomposed into two tractable
sub-problems: maximizing the successful transmission rate of images
(MSTRI) and ensuring the timeliness problem (ETP). Subsequently, an
efficient subgradient of relaxation computing constraint (SRCC) algorithm
is proposed. The upper and lower bounds of MSTR-TDI are obtained by
solving the two subproblems and the dual problem (DP), and the direction
of the next iteration is obtained by feedback. Furthermore, arranging
the sending sequences of images to improve the quality of the solution.
The approximate optimal solution of MSTR-TDI is eventually obtained
through repeated iterations. The simulation results verify the superiority
of the proposed MDR-TEG model and the effectiveness of the\textcolor{red}{{}
}SRCC.

\end{abstract}

\section{Introduction\label{sec:Introduction}}

\IEEEPARstart{W}{}ith the development of space-ground integrated
network (SGIN), the low earth orbit satellites (LEO), with their advantages
of wide coverage, low transmission delay, and frequency reuse efficiency,
have become an important role in the sixth generation (6G) mobile
communication system\cite{Zhang2020,Sheng2019,Jia2021LOT,9693471}.
Furthermore, the traffic of time-deterministic remote sensing image
in networks has shown explosive growth\cite{Wang2022}. According
to Cisco's white paper on the visual networking index, global mobile
data traffic in 2022 increased by six times compared to 2017, of which
video/image traffic accounted for 82\%, and live video/image traffic
was 16 times higher than in 2017 \cite{example2022}. These images
involve considerable traffic volume and strict end-to-end delay requirements.
The efficient and collaborative use of scarce resources to ensure
time-deterministic image transmission has become an urgent issue.
To tackle this issue, research on onboard computing technologies has
increasingly become a popular field for ensuring the quality of service
(QoS) \cite{KaibinHuang2017survey,Wangshuo}. However, due to the
heterogeneity and differences in the communication, caching, and computing
(3C) resources of bearing satellites, the lack of low-complexity methods
for characterization and effective collaborative allocation strategies
has resulted in a low successful transmission ratio, further making
the issue even more challenging \cite{Kaibinhuang_effective_allocation,HayderAl-Hraishawi}.

\subsection{Related Work\label{subsec:1.1 Related Work}}

\begin{figure}[t]
\begin{centering}
\includegraphics{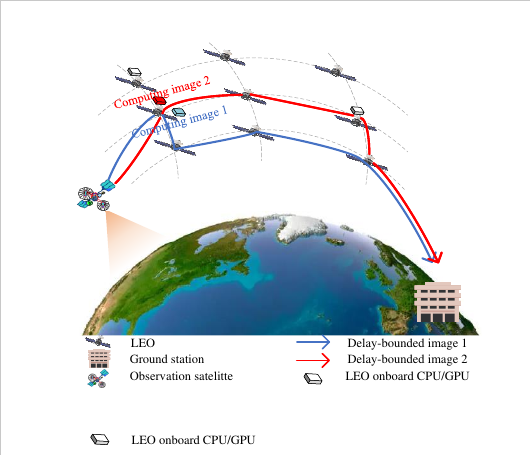}
\par\end{centering}
\centering{}\caption{Multi-dimensional Resource Services Image Applications in the SGIN\label{fig1:Scene of SGIN}}
\end{figure}

The traditional sensing service transmission method is that the observation
satellite (OS) sends the service to the data processing center through
the LEO network for calculation and analysis, and then returns the
control information to the OS \cite{Jia2021}. The long delay cannot
meet the transmission requirements of the service. With the development
of hardware technology, the onboard computing functions of satellites
are realized. As shown in Fig. \ref{fig1:Scene of SGIN}, the image/video
collected by the OS is uploaded to the LEO satellite first. Then the
business undergoes the process of storage, calculation(i.e., the processing
of image and video), and transmission. Finally transmitted to the
ground station (GS) through the satellite-ground link, and then delivered
by the GS to the data processing center for analysis. In this procedure,
the sensing application requires the 3C resources of the network.

\begin{figure}[t]
\begin{centering}
\includegraphics[scale=0.47]{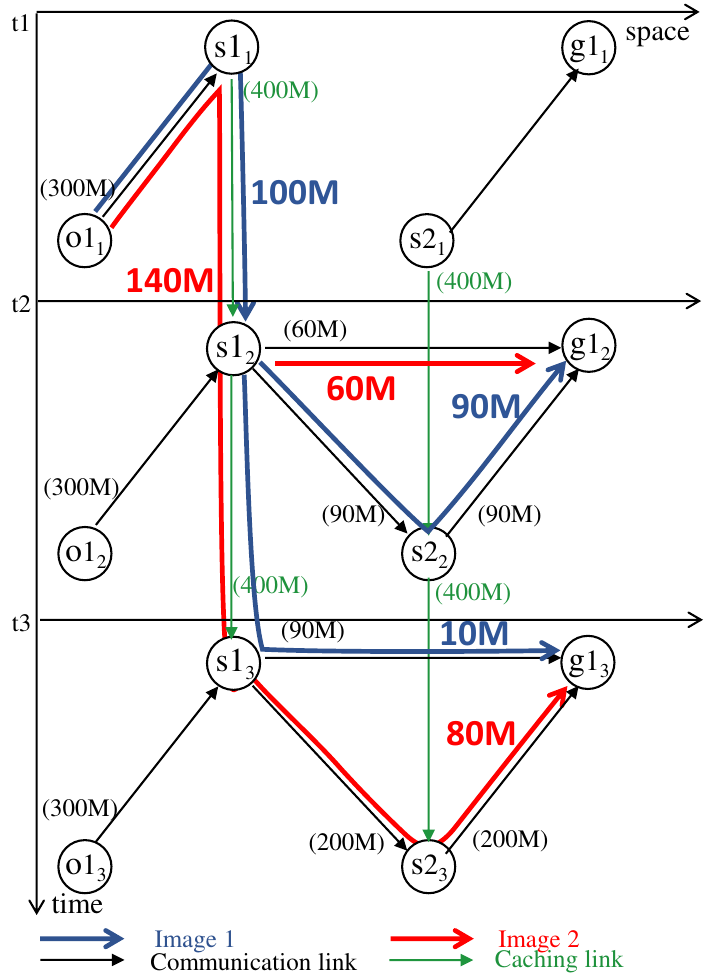}
\par\end{centering}
\centering{}\caption{Examples of resource characterization and path planning based on TEG.\label{fig:Examples-of-resource-characterization}}
\end{figure}

To make full use of network resources, some researchers focus on the
reasonable characterization of which. At present, the modeling is
usually based on the time-expanded graph (TEG)\cite{Zhou2018,Konidaris2013}.
However, TEG can only represent two-dimensional resources (i.e. storage
and transmission resources) in the network and cannot characterize
other resources\cite{Yang2022}. Therefore, the image delay requirement
cannot be met. Assuming that there are two images with delay boundaries
of 2 slots that need to be sent from the source $o1$ to the destination
$g$. The resource representation based on TEG is shown in Fig. \ref{fig:Examples-of-resource-characterization}.
Since computing resources cannot be represented, only the storage
and transmission resources already represented in the SGIN can be
allocated. When $V_{s_{1}}$ has no $\varepsilon_{ss}$ in slot 1,
it can only store service to the next time slot. Due to the limitation
of transmission resources, at least 3 slots are required to complete
the transmission. And the delay of the two images will exceed the
boundaries. Therefore, some scholars have made improvements based
on TEG. Wang et al. \cite{Wang2019} divided the OS into the image
acquisition part and compression calculation part according to function,
and modeled the problem as a random network optimization problem with
network utility maximization. The author in \cite{Wang2018} characterized
the computing capacity of the destination node by establishing the
connection between the destination node and the virtual node and modeled
the joint allocation of resources as a maximum flow problem. Liu et
al. \cite{Liu2023} designed a multi-functional time expansion graph
(MF-TEG). Each node generates multiple copies to represent different
computing functions and verifies that rational utilization of computing
resources can increase the total traffic of the network. Nevertheless,
generating a copy of each node increases the complexity of graph-based
representation and the difficulty of solving it. 

And there have been many great studies on the joint optimization of
resources allocation \cite{Zhou2019,Du2018,PengwenlongGu,HayderAl-Hraishawi,Du2017}.
The optimization objectives are divided into three categories, and
the objective functions are usually end-to-end delay minimization
\cite{Shi2020,DaiPenglin2020}, energy minimization \cite{Li2024,KimJunghoon2020,WangKezhi2018,7727952}
and computation rate maximization \cite{Li2023}. An interesting work
is to explore the influence of computing resources on delay. Ren et
al.\cite{Ren2018} explored that partial or full calculation of acquired
images can significantly improve the delay performance. In \cite{Ren2019},
they further study the network delay with different computing locations
and point out that the order of multi-dimensional resource allocation
will affect the network performance.

The previous works provide invaluable insights for our research. As
far as the authors know, the above researchers often take the computational
capabilities of ground stations or source satellites into account,
with relatively less consideration given to the LEO satellites in
a bearing network of the SGIN. However, differing from the ground
network, the LEO exhibits high mobility, resulting in a dynamic yet
predictable topology. The inter-satellite links (ISL) will also be
on and off with the movement, and the bandwidth is severely limited.
There is a lack of efficient descriptions of the LEO satellites. Therefore,
characterizing the resources of relay satellites is also equally important.
Furthermore, facing the latency requirements of huge-volume images,
the prompt allocation of precious resources on board is even more
urgent. Meanwhile, even if the computing resources are utilized, if
the path is not adjusted in time to match the image traffic (although
the traffic of data of the compressed image is reduced, it is still
transmitted along the original path), image 2 will eventually time
out.

To solve the above problems, in this paper, we mainly focus on the
low-complexity representation model based on TEG and efficient resource
allocation algorithm design to enrich the theory and practice in the
field of on-board computing research. And the comparison between this
job and others are shown as Table \ref{tab:A COMPARISON BETWEEN THIS JOB AND OTHERS.}.

\subsection{Contribution\label{subsec:1.2 Contribution}}

To solve these problems, a multi-dimensional resource representation
model is proposed and an efficient algorithm is designed. The contributions
of this paper are summarized as fourfold aspects:
\begin{itemize}
\item To demonstrate computing resources distinctly, we propose a multi-dimensional
resource time-expanded graph (MDR-TEG) model. Firstly, the storage
and transmission resources are characterized based on TEG. Secondly,
all computing resources in the SGIN are centrally represented as a
virtual node to reduce the number of computing nodes, and arcs are
established between nodes and virtual nodes to represent the computing
process. Finally, a virtual absorption node is established to capture
the computing loss. Since only two virtual nodes are added, the complexity
of resource representation is greatly reduced.
\item The problem is modeled as maximizing the successful transmission ratio
of time-deterministic images (MSTR-TDI) based on the proposed MDR-TEG.
The MSTR-TDI is a mixed integer linear programming (MILP) problem,
whose nondeterministic polynomial hard (NP-hard) characteristics make
it impractical to solve outright. Therefore, we further transform
the MSTR-TDI into the Lagrange dual problem (DP) by introducing Lagrange
multipliers to release the difficult constraints. The mathematical
derivations proved that the optimal solution of DP is the upper bound
of the original problem, which is tighter than that of linear relaxation.
\item The original problem is formulated as maximizing the successful transmission
ratio of time-deterministic images (MSTR-TDI) using the proposed MDR-TEG
model. This MSTR-TDI problem constitutes a mixed integer linear programming
(MILP) formulation and exhibits nondeterministic polynomial-time hard
(NP-hard) characteristics, rendering direct solution intractable.
To address this computational challenge, we introduce Lagrange multipliers
to relax the complicating constraints, thereby decomposing the MSTR-TDI
into two more tractable subproblems. To efficiently coordinate the
solutions of these subproblems and progressively approach the optimal
solution, we leverage a subgradient of relaxation computing constraint
(SRCC) framework. This algorithm iteratively refines the solution
by updating the multipliers to narrow the gap between the upper and
lower bounds, ensuring a monotonically improving sequence of feasible
solutions.
\item The simulation results show that the execution time of the proposed
algorithm is only 0.01\% of the exhaustive search algorithm based
on Lagrange relaxation (ESALR), with a successful transmission ratio
error of just 8.57\%. Compared to other contrast algorithms (i.e.,
joint allocation (JA) algorithm and computing resources priority allocation
algorithm (CRPAA)), the proposed SRCC has significantly increased
the performance in running time, end-to-end delay, and transmission
success ratio.
\end{itemize}
The follow-up of this paper is arranged as follows. Section II presents
the multi-dimensional resource time-expanded graph (MDR-TEG) model,
introducing the capacity constraints (i.e., transmission capacity,
caching capacity, and computational constraints), flow conservation,
compression constraint and delay constraint. The problem formulation
and transformation are proposed in Section III. Section IV designs
the subgradient of relaxation computing constraint (SRCC) algorithm.
The simulation results are shown in section V. And section VI is the
conclusion of our work.%

\section{System Model \label{sec2:System Model}}

The multi-dimensional resource time-expanded graph (MDR-TEG) $\mathbb{\mathbb{M}}=\left\{ V,\varepsilon,T\right\} $
is used to analyze the SGIN. We divide the continuous time range $[0,T)$
into K small time slots, and let $\tau=T/K$ denote the slot length.
$[t-1,t)$ is considered the $t$th slot. Assuming that the satellites\textquoteright{}
position is stationary during the time slot $\tau$ and changes between
two slots. The directed graph represents the MDR-TEG, where:
\begin{itemize}
\item Denote the vertices $V=\{V_{o}\cup V_{s}\cup V_{g}\cup V_{com}\cup V_{a}\}$
as the set of all nodes in the network, in which $V_{o}$ is the set
of OSs, $V_{s}$ is the set of LEO satellites, and $V_{g}$ represents
the set of the GSs. $V_{com}$ is a virtual computing node, and $V_{a}$
is the virtual absorption point, representing the reduced data volume
due to the computing during time $T$.
\item Denote the links $\varepsilon=\{\varepsilon_{os}\cup\varepsilon_{ss}\cup\varepsilon_{sg}\cup\varepsilon_{sc}\cup\varepsilon_{cs}\cup\varepsilon_{ca}\cup\varepsilon_{s}\}$
as the set of all links in the network, where $\varepsilon_{t}=\{\varepsilon_{os}\cup\varepsilon_{ss}\cup\varepsilon_{sg}\}$
are the transmission links, the $\varepsilon_{s}$ represents the
caching links that can store data between the slot $t$ to $t+1$.
The $\varepsilon_{sc}$ is the computing links and $\varepsilon_{cs}$
is the computed return links, and the $\varepsilon_{ca}$ is the absorption
links. 
\end{itemize}
Based on TEG, MDR-TEG adds $V_{com}$ and $V_{a}$, and the computing
function of the node is considered as a potential link (i.e., the
gray link) between the node and $V_{com}$. The link weight between
each LEO and $V_{com}$ can be regarded as a mapping of the computing
capacity of the LEO. $V_{s}$ has only transfer and storage capabilities,
and the computation can be viewed as generating an arc from $V_{s}$
to $V_{com}$. Then the network can allocate three-dimensional resources.
As shown in Fig. \ref{fig:MDR-TEG(b)-new-1}, The image 1 will go
through the following transfer process: The $V_{o_{1}}$ firstly sends
the observation result image 1 and image 2 to the $V_{s_{1}}$ through
the link $\varepsilon_{os}$, then $V_{s_{1}}$ computing it using
the link $\varepsilon_{sc}$. After that, the data loss is sent to
the $V_{a}$ through the link $\varepsilon_{ca}$, while the computed
result returns to $V_{s_{1}}$. Due to compression decreasing the
data volume, the mission could eventually takes 2 slots to reach the
destination via the link $\varepsilon_{sg}$.

\subsection{Capacity Constraint\label{subsec:2.1 Capacity Constraint}}

\subsubsection{Transmission capacity constraint}

The connection between satellites is constantly changing due to the
high-speed movement. This leads to the dynamic change of ISL rate,
assuming that the connection relationship between nodes only changes
in the time slot, and is static in the time slot, so we can know the
link rate $c(i_{t},j_{t}),(i_{t},j_{t})\in\{\varepsilon_{os}\cup\varepsilon_{ss}\}$
between two satellites before the start of the time slot, which can
be obtained by the following formula \cite{9261433}:
\begin{figure}[t]
\centering{}\includegraphics[scale=0.65]{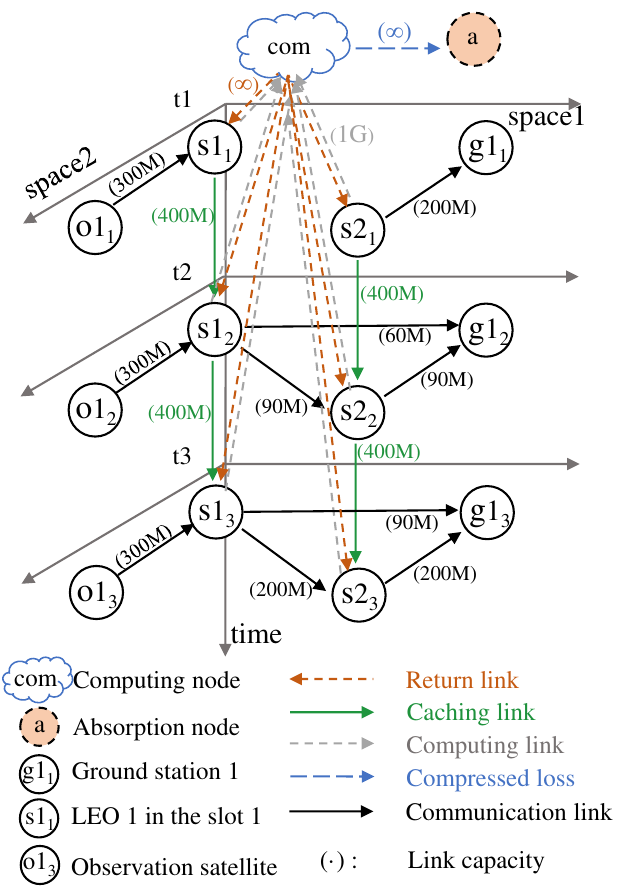}\caption{Multi-dimensional resource representation based on MDR-TEG and adjust
the path in time.\label{fig:MDR-TEG(b)-new-1}}
\end{figure}
\begin{table*}[t]
\centering{}\caption{\label{tab:A COMPARISON BETWEEN THIS JOB AND OTHERS.}A COMPARISON
BETWEEN THIS JOB AND OTHERS.}
\begin{tabular}{cccccc}
\hline 
\multirow{2}{*}{\textbf{Research}} & \textbf{The OS} & \textbf{The Leo} & \textbf{The GS} & \multirow{2}{*}{\textbf{Multiple Virtual Nodes}} & \multirow{2}{*}{\textbf{A Virtual Node}}\tabularnewline
 & \textbf{Computing} & \textbf{Computing} & \textbf{Computing} &  & \tabularnewline
\hline 
 &  &  &  &  & \tabularnewline
Liu et al.\cite{LiuRunzi2016} & $\checked$ &  &  & $\checked$ & \tabularnewline
Yang et al.\cite{Liu2023} &  & $\checked$ &  & $\checked$ & \tabularnewline
Wang et al.\cite{Wang2018} &  &  & $\checked$ & $\checked$ & \tabularnewline
Our Work & $\checked$ & $\checked$ &  &  & $\checked$\tabularnewline
\hline 
\end{tabular}
\end{table*}

\begin{equation}
c(i_{t},j_{t})=\frac{P_{tr}G_{re}G_{tr}L_{pl}(t)\gamma_{s}}{k_{B}T_{s}(E_{b}/N_{0})_{req}M}
\end{equation}

The transmission power of $i_{t}$ is defined as $P_{tr}$ (in W),
the $G_{re}$ and $G_{tr}$ are the receiving and transmitting antenna
gain of $j_{t}$ and $i_{t}$ (in dB), respectively. The $L_{pl}$
represents the free space loss (in dB), and $\gamma_{s}$ is the total
line loss (in dB). $k_{B}$ is Boltzmann\textquoteright s constant
(in $JK^{-1}$), $T_{s}$ is total system noise temperature (in K),
$(E_{b}/N_{0})_{req}$ and $M$ are the required ratio of received
energy-per-bit to noise-density and link margin, respectively. $L_{l}$
can be obtained from the formula $L_{pl}(t)=\left(c/4\pi d(t)f_{s}\right)^{2}$,
where $c$ is the speed of light (in km/s), $d(t)$ is the maximum
slant range (in km), and $f_{s}$ is the communication center frequency
(in Hz).

According to\textcolor{magenta}{{} }\cite{Zhu2020,7581088}, the link
rate $c(i_{t},j_{t}),(i_{t},j_{t})\in\{\varepsilon_{sg}\}$ between
satellites and ground stations can be obtained by the following equation

\begin{equation}
c(i_{t},j_{t})=B\log_{2}\left(1+SNR(i_{t},j_{t})\right)
\end{equation}
where $B$ indicates the channel bandwidth (in Hz), and $SNR$ is
the signal-to-noise ratio (in dB). Besides, SNR can be expressed as

\begin{equation}
SNR(i_{t},j_{t})=\left(\frac{P_{tr}G_{tr}(t)G_{re}L_{pl}(t)\gamma_{r}(t)}{n_{0}B}\right)^{2}
\end{equation}
$P_{tr}$ (in W) and $G_{tr}(t)$ (in dB) represent the satellite's
transmit power and transmit antenna gain, respectively. $G_{re}$
is the receive antenna gain of the ground station (in dB). $L_{pl}(t)$
stands for the satellite-ground link path loss (in dB) which can be
calculated from (8), and $\gamma_{r}(t)$ is the rain attenuation
(in dB). $n_{0}$ is the noise power density (in W/Hz), $n_{0}B$
is the noise of the channel (in dB). During the slot $t$, the transmission
link capacity is equal to the product of the transmission link rate
and slot length:

\begin{equation}
C(i_{t},j_{t})=c(i_{t},j_{t})\cdot\tau,\forall(i_{t},j_{t})\in\varepsilon_{t},t\in T
\end{equation}
We further denote the images as the set of flows $F$ in the SGIN,
the arbitrary $u\in F$ is a five-tuple, i.e., $u=\{s,d,\delta_{u},T_{u}^{s},T_{u}^{e}\}$.
Where $s$ is the source node, $d$ is the destination node, $\delta_{u}$
(in bits) is the volume of the flow $u$, $T_{u}^{s}$ and $T_{u}^{e}$
are the start and end transmission time respectively. The $x_{u}(i_{t},j_{t})$
represents the amount of $u$ transmitted on the link $(i_{t},j_{t})$
(in bits), and the resources consumption matrix $x=\{x_{u}(i_{t},j_{t})|\forall(i_{t},j_{t})\in\varepsilon,u\in F,t\in T\}$
contains the usage information of the resources by $F$ in $T$. In
slot $t$, all flows passing through the link $(i_{t},j_{t})$ must
not exceed the upper bound of the transmission link capacity:

\begin{equation}
\underset{u\in F}{\sum}x_{u}(i_{t},j_{t})\leq C(i_{t},j_{t}),\forall(i_{t},j_{t})\in\varepsilon_{t},t\in T
\end{equation}

\subsubsection{Caching capacity constraint}

Depending on the characteristics of the memory hardware, all flows
passing through the caching link $(i_{t},i_{t+1})$ must not exceed
the maximum number of bits $S_{max}$ (in bits) of the on-board memory:

\begin{equation}
\underset{u\in F}{\sum}x_{u}(i_{t},i_{t+1})\leq S_{max},\forall i_{t}\in V_{s},1\leq t<T
\end{equation}

In particular, in order to ensure the rationality, before the first
slot, and after the last slot, the capacity of the storage link is
0.

\subsubsection{Computing capacity constraint}

In slot $t$, all flows passing through the link $(i_{t},V_{com})$
must not exceed the computing link capacity:

\begin{equation}
\underset{u\in F}{\sum}x_{u}(i_{t},V_{com})\leq\rho\cdot\zeta_{max},\forall i_{t}\in V_{s},t\in T\label{eq:2.1.3 Computing capacity constraint}
\end{equation}
Where $\rho$ (in Mbits/units) and $\zeta_{max}$ (in units) reflects
the computing capacity of the CPU on the satellite. Furthermore, the
capacity of virtual link $\varepsilon_{cl}$ can be considered infinite,
i.e. $C(V_{com},V_{l})=\infty$.

\subsection{Flow Conservation\label{subsec:2.2 Flow Conservation}}

\subsubsection{The flow conservation of source node}

The binary variable $\lambda_{u}(i_{t},j_{t})$ is defined primarily
to represent whether flow $u$ passes the link $(i_{t},j_{t})$:

\begin{equation}
\lambda_{u}(i_{t},j_{t})=\begin{cases}
1, & \textrm{if \ensuremath{x_{u}}(\ensuremath{i_{t}},\ensuremath{j_{t}}) > 0},(i_{t},j_{t})\in\varepsilon\\
0, & \textrm{otherwise}
\end{cases}\label{eq:x and lamuda}
\end{equation}
The path choose matrix is $\lambda=\{\lambda_{u}(i_{t},j_{t})|\forall(i_{t},j_{t})\in\varepsilon,u\in F,t\in T\}$.
The source node can select only one link to send flow $u$:

\begin{equation}
\underset{(i_{t},j_{t})\in\varepsilon_{os}}{\sum}\lambda_{u}(i_{t},j_{t})=1,\forall i_{t}\in V_{o},u\in F,t\in T
\end{equation}

The amount of outgoing data equals the volume of the of flow $u$:

\begin{equation}
\underset{(i_{t},j_{t})\in\varepsilon_{os}}{\sum}x_{u}(i_{t},j_{t})=\delta_{u},\forall i_{t}\in V_{o},u\in F,t\in T
\end{equation}

\subsubsection{The flow conservation of destination node}

The destination node can also select only one link to receive flow
$u$:

\begin{equation}
\begin{array}{l}
\underset{(i_{t},j_{t})\in\varepsilon_{sg}}{\sum}\lambda_{u}(i_{t},j_{t})=1,\forall i_{t}\in V_{s},u\in F,t\in T\end{array}
\end{equation}

\subsubsection{The flow conservation of ordinary node}

Towards any LEO node $j_{t}$ in the network, its compression function
can be separated from the ontology communication and storage function,
and integrated into the virtual node $V_{com}$. Therefore, the LEO
satellite vertice can be regarded as an ordinary node with only storage
and transmission functions. Furthermore, when the stream $u$ has
computing requirements, the compression procedure can be represented
by enabling link $(j_{t},V_{com})$. The processed result of $u$
is returned to $j_{t}$ along the link $(V_{com},j_{t})$. It can
be seen from Fig. \ref{fig:MDR-TEG(b)-new-1} that there are three
kinds of links flowing into $j_{t}$ : the transmission link $(i_{t},j_{t})$,
the storage link $(j_{t-1},j_{t})$ and the calculation link $(j_{t},V_{com})$
. \textcolor{brown}{}

The flow $u$ can only enter $j_{t}$ from one actual node and output
to a physical node, set the $\lambda_{left}=\underset{(i_{t},j_{t})\in\varepsilon_{os}\cup\varepsilon_{ss}}{\sum}\lambda_{u}(i_{t},j_{t})+\lambda_{u}(j_{t-1},j_{t})$
and the $\lambda_{right}=\underset{(j_{t},i_{t})\in\varepsilon_{ss}\cup\varepsilon_{sg}}{\sum}\lambda_{u}(j_{t},i_{t})+\lambda_{u}(j_{t},j_{t+1})$.
Then:

\begin{equation}
\lambda_{left}=\lambda_{right}=1,\forall j_{t}\in V_{s},u\in F,t\in T
\end{equation}

Set the $x_{left}=\underset{(i_{t},j_{t})\in\varepsilon_{os}\cup\varepsilon_{ss}\cup\varepsilon_{cs}}{\sum}x_{u}(i_{t},j_{t})+x_{u}(j_{t-1},j_{t})+x_{u}(V_{com},V_{a})$,
and $x_{right}=\underset{(j_{t},i_{t})\in\varepsilon_{ss}\cup\varepsilon_{sg}\cup\varepsilon_{sc}}{\sum}x_{u}(j_{t},i_{t})+x_{u}(j_{t},j_{t+1})$.\textcolor{magenta}{{}
}The volume of flow $u$ must satisfied:

\begin{equation}
x_{left}=x_{right},\forall j_{t}\in V_{s},u\in F,t\in T
\end{equation}

\subsubsection{The flow conservation of virtual computing node}

If $\lambda_{u}(i_{t},V_{com})=1$, it represents that flow $u$ is
compressed at node $i$ in slot $t$. The flow volume should be equal
to:

\begin{equation}
\begin{array}{l}
x_{u}(i_{t},V_{com})\cdot\lambda_{u}(i_{t},V_{com})=\underset{(j_{t},i_{t})\in\varepsilon_{os}\cup\varepsilon_{ss}}{\sum}x_{u}(j_{t},i_{t})\\
+x_{u}(i_{t-1},i_{t}),\forall i_{t}\in V_{s},u\in F,t\in T
\end{array}\label{eq:Vcom constraint}
\end{equation}

The amount of computing result returned $x_{u}(V_{com},i_{t})$ and
compress losing $x_{u}(V_{com},V_{a})$ are related to the $x_{u}(i_{t},V_{com})$:

\begin{equation}
\begin{array}{l}
x_{u}(i_{t},V_{com})=x_{u}(V_{com},i_{t})+x_{u}(V_{com},V_{a})\\
,\forall i_{t}\in V_{s},u\in F,t\in T
\end{array}
\end{equation}

The computing loss satisfies $x_{u}(V_{com},V_{a})=x_{u}(i_{t},V_{com})\cdot(1-\theta_{u})$,
Where $\theta_{u}\in(0,1)$ is compression ratio of flow $u$. Meaning
that the size of the $u$ returned after compression is $\theta_{u}$
times that of the flow before compression.

\subsection{Compression constraint\label{subsec:2.3 Compression constraint}}

Considering the real scenario (i.e., the image can only be compressed
once at most \cite{Liu2023a}), the flow compression constraint is
given, and the stream $u$ can only flow into $V_{com}$ once at most:

\begin{equation}
\underset{t\in T}{\sum}\underset{i_{t}\in V_{s}}{\sum}\lambda_{u}(i_{t},V_{com})\leq1,\forall u\in F\label{eq:2.3 Compression constraint}
\end{equation}

\subsection{Delay constraint\label{subsec:2.4 Delay constraint}}

For $\forall j_{t}\in V_{g},u\in F,t\in T$, the flow $u$ which is
successfully transmitted to the destination, its arrival time is recorded:

\begin{equation}
T_{u}=\begin{cases}
t, & \textrm{if \ensuremath{\lambda_{u}}(\ensuremath{i_{t}},\ensuremath{j_{t}})=1}\\
-1, & \textrm{otherwise}.
\end{cases}\label{eq:lamda and Tu}
\end{equation}

The binary variable $\lambda_{u}^{sg}(i_{t},j_{t})$ is defined to
represent whether flow $u$ successfully reaches the destination within
the delay limit, which is a part of the timely transmission matrix
$\lambda^{sg}=\{\lambda_{u}^{sg}(i_{t},j_{t})|\forall(i_{t},j_{t})\in\varepsilon,u\in F,t\in T\}$,
and:

\begin{equation}
\lambda_{u}^{sg}(i_{t},j_{t})=\begin{cases}
1, & \textrm{if }T_{u}^{s}\leq T_{u}\leq T_{u}^{e}\\
0, & \textrm{otherwise}.
\end{cases}\label{eq:Tu and lamda_GS}
\end{equation}

\section{The Problem Formulation And Decomposition}

\subsection{Problem Formulation\label{subsec:3.1 Problem Formulation}}

In this paper, our optimization goal is defined as maximizing the
successful transmission ratio of time-deterministic images (MSTR-TDI),
the problem is formulated as:

\begin{align*}
\mathbf{MSTR-TDI}:\underset{\left\{ \lambda,x,T_{u},\lambda_{u}^{sg}\right\} }{max} & \underset{t\in T}{\sum}\underset{u\in F}{\sum}\underset{j_{t}\in V_{g}}{\sum}\lambda_{u}^{sg}(i_{t},j_{t})\\
s.t. & (5)-(18)
\end{align*}

The nonlinear constraints (\ref{eq:x and lamuda}), (\ref{eq:Vcom constraint}),
(\ref{eq:lamda and Tu}) , (\ref{eq:Tu and lamda_GS}) and discrete
variables make MSTR-TDI nonconvex and complicated \cite{Boyd2024ConvexOptimization}.
It is necessary to linearize them and reduce the variables.

\subsection{Problem Transformation \label{subsec:3.2 Problem Transformation-1}}

Firstly, We introduce a large positive integer M, the equation (\ref{eq:x and lamuda})
is converted to:

\begin{equation}
\lambda_{u}(i_{t},j_{t})\leq\ensuremath{x_{u}}(\ensuremath{i_{t}},\ensuremath{j_{t}})\leq M\cdot\lambda_{u}(i_{t},j_{t}),(i_{t},j_{t})\in\varepsilon\label{eq:x and lamuda- - linked}
\end{equation}

when $\lambda_{u}(i_{t},j_{t})=0$, $\ensuremath{x_{u}}(\ensuremath{i_{t}},\ensuremath{j_{t}})=0$,
and $\lambda_{u}(i_{t},j_{t})=1$, $\ensuremath{x_{u}}(\ensuremath{i_{t}},\ensuremath{j_{t}})$
is a positive integer.\textcolor{magenta}{{} }Let $\underset{(j_{t},i_{t})\in\varepsilon_{os}\cup\varepsilon_{ss}}{\sum}x_{u}(j_{t},i_{t})+x_{u}(i_{t-1},i_{t})=z$,
and (\ref{eq:Vcom constraint}) can be rewritten as:

\begin{equation}
\begin{array}{l}
x_{u}(i_{t},V_{com})-M\cdot(1-\lambda_{u}(i_{t},V_{com}))\leq z\leq x_{u}(i_{t},V_{com})\\
0\leq z\leq M\cdot\lambda_{u}(i_{t},V_{com})\\
,\forall i_{t}\in V_{s},u\in F,t\in T
\end{array}\label{eq:Vcom constraint- -linked}
\end{equation}

Similarly, equations (\ref{eq:lamda and Tu}) and (\ref{eq:Tu and lamda_GS})
can be expressed as:

\begin{equation}
t=(t-M)\cdot\lambda_{u}(i_{t},j_{t})+M,\forall j_{t}\in V_{g},u\in F,t\in T\label{eq:lamda and t- -linked}
\end{equation}

\begin{equation}
\begin{array}{lc}
t\geq T_{u}^{s}-M\cdot[1-\lambda_{u}^{sg}(i_{t},j_{t})]\\
t\leq T_{u}^{e}+M\cdot[1-\lambda_{u}^{sg}(i_{t},j_{t})] & ,\forall j_{t}\in V_{g},u\in F,t\in T\\
t\geq T_{u}^{e}+1-M\cdot\lambda_{u}^{sg}(i_{t},j_{t})
\end{array}\label{eq:lamda-GS and t- -linked}
\end{equation}

Then, put the (\ref{eq:lamda and t- -linked}) into the (\ref{eq:lamda-GS and t- -linked}),
we finally obtain the linear constraints on $t$ and $\lambda_{u}^{sg}(i_{t},j_{t})$:

\begin{equation}
\begin{array}{l}
\lambda_{u}^{sg}(i_{t},j_{t})\leq\dfrac{(t-M)}{M}\cdot\lambda_{u}(i_{t},j_{t})+\dfrac{(2M-T_{u}^{s})}{M}\\
\lambda_{u}^{sg}(i_{t},j_{t})\leq\dfrac{(M-t)}{M}\cdot\lambda_{u}(i_{t},j_{t})+\dfrac{T_{u}^{e}}{M}\\
\lambda_{u}^{sg}(i_{t},j_{t})\geq\dfrac{(M-t)}{M}\cdot\lambda_{u}(i_{t},j_{t})+\dfrac{(1+T_{u}^{e})-M}{M}\\
,\forall u\in F,t\in T
\end{array}\label{eq:lamda-GS and t- -linked-1}
\end{equation}

Finally, the MSTR-TDI is equivalently transformed to the following
standard form:

\begin{align*}
\mathbf{OP}:\underset{\left\{ \lambda,x,\lambda_{u}^{sg}\right\} }{min} & -\underset{t\in T}{\sum}\underset{u\in F}{\sum}\underset{j_{t}\in V_{g}}{\sum}\lambda_{u}^{sg}(i_{t},j_{t})\\
s.t.\textrm{C1}: & (5)-(7)\\
\textrm{C2}: & (9)-(13)\text{,}(15)-(16),(19)-(20)\\
\textrm{C3}: & (23)
\end{align*}

In the OP, C1 is the capacity constraint, C2 is the flow conservation
constraint and C3 is the delay constraints. It can be seen that this
problem is a mixed integer linear programming (MILP) problem, which
is a non-deterministic polynomial hard (NP-hard) problem obviously.
The solution time of OP is related to the size of variables and the
number of constraints in the network. Therefore, it is necessary to
design an efficient solution algorithm.

\subsection{Problem Decomposition \label{subsec:3.3 Problem Decomposition}}

Since the time to directly solve OP is inestimable. It can be seen
that formula (\ref{eq:lamda-GS and t- -linked-1}) couples $\lambda_{u}(i_{t},j_{t})$
and $\lambda_{u}^{sg}(i_{t},j_{t})$ together and makes the problem
difficult. To circumvent these hurdles, we introduce the Lagrange
multiplier method to obtain the solution. what's more, the solution
obtained by Lagrangian relaxation is usually tighter than that obtained
by linear relaxation \cite{Sun2010}. Therefore, let $g_{1}(\lambda,x,\lambda_{u}^{sg})=\lambda_{u}^{sg}(i_{t},j_{t})-\dfrac{(t-M)}{M}\cdot\lambda_{u}(i_{t},j_{t})-\dfrac{(2M-T_{u}^{s})}{M}\leq0$
, $g_{2}(\lambda,x,\lambda_{u}^{sg})=\lambda_{u}^{sg}(i_{t},j_{t})-\dfrac{(M-t)}{M}\cdot\lambda_{u}(i_{t},j_{t})-\dfrac{T_{u}^{e}}{M}\leq0$
and $g_{3}(\lambda,x,\lambda_{u}^{sg})=-\lambda_{u}^{sg}(i_{t},j_{t})+\dfrac{(M-t)}{M}\cdot\lambda_{u}(i_{t},j_{t})+\dfrac{(T_{u}^{e}+1)-M}{M}\leq0$.
we fist introduce multipliers $\mu_{1}^{i,u,t}$, $\mu_{2}^{i,u,t}$
and $\mu_{3}^{i,u,t}\geq0$ to get the Lagrange function $\boldsymbol{L}(\lambda,x,\lambda_{u}^{sg},\mu_{1}^{i,u,t},\mu_{2}^{i,u,t},\mu_{3}^{i,u,t})$:

\begin{equation}
\begin{array}{c}
\begin{array}{c}
\begin{array}{l}
\boldsymbol{L}(\lambda,x,\lambda_{u}^{sg},\mu_{1}^{i,u,t},\mu_{2}^{i,u,t},\mu_{3}^{i,u,t})\\
=-\underset{t\in T}{\sum}\underset{u\in F}{\sum}\underset{j_{t}\in V_{g}}{\sum}\lambda_{u}^{sg}(i_{t},j_{t})\\
+\underset{t\in T}{\sum}\underset{u\in F}{\sum}\underset{j_{t}\in V_{g}}{\sum}\mu_{1}^{i,u,t}\cdot g_{1}(\lambda,x,\lambda_{u}^{sg})\\
+\underset{t\in T}{\sum}\underset{u\in F}{\sum}\underset{j_{t}\in V_{g}}{\sum}\mu_{2}^{i,u,t}\cdot g_{2}(\lambda,x,\lambda_{u}^{sg})\\
+\underset{t\in T}{\sum}\underset{u\in F}{\sum}\underset{j_{t}\in V_{g}}{\sum}\mu_{3}^{i,u,t}\cdot g_{3}(\lambda,x,\lambda_{u}^{sg})
\end{array}\end{array}\end{array}\label{eq:Lagrange function}
\end{equation}

Further, we propose the Lagrange dual problem (DP) of the OP:

\begin{align*}
\mathbf{DP}: & \underset{\left\{ \mu_{1}^{i,u,t},\mu_{2}^{i,u,t},\mu_{3}^{i,u,t}\right\} }{max}d(\mu)\triangleq\\
 & \underset{\left\{ \mu_{1}^{i,u,t},\mu_{2}^{i,u,t},\mu_{3}^{i,u,t}\right\} }{max}\underset{\left\{ \lambda,x,\lambda_{u}^{sg}\right\} }{min}\boldsymbol{L}(\lambda,x,\lambda_{u}^{sg},\mu_{1}^{i,u,t},\mu_{2}^{i,u,t},\mu_{3}^{i,u,t})
\end{align*}

\begin{proposition} 

Assume that the optimal solution of the dual problem is $d^{*}$ and
the optimal solution of the OP $f(\lambda,x,\lambda_{u}^{sg})$ is
$p^{*}$. According to the weak duality theorem, the $d^{*}$ is the
maximum lower bound of the $p^{*}$ (for minimization problems).

\label{proposition: lower bound}

\end{proposition}

\textit{Proof: }Suppose $\stackrel{\:\bullet}{\lambda}$, $\stackrel{\:\bullet}{x}$,
and $\stackrel{\:\bullet}{\lambda_{u}^{sg}}$ are arbitrary points
of the feasible domains \textit{$\mathcal{D}_{1}$,$\mathcal{D}_{2}$
and $\mathcal{D}_{3}$}, respectively. The $d(\mu)$ can be expanded
as:

\begin{equation}
\begin{array}{rl}
d(\mu) & =\underset{\{\stackrel{\:\bullet}{\lambda}\in\mathcal{D}_{1},\stackrel{\:\bullet}{x}\in\mathcal{D}_{2},\stackrel{\:\bullet}{\lambda_{u}^{sg}}\in\mathcal{D}_{3}\}}{inf}\boldsymbol{L}(\stackrel{\:\bullet}{\lambda},\stackrel{\:\bullet}{x},\stackrel{\:\bullet}{\lambda_{u}^{sg}},\mu_{1}^{i,u,t},\mu_{2}^{i,u,t},\mu_{3}^{i,u,t})\\
 & =f(\mathit{\stackrel{\:\bullet}{\lambda},\stackrel{\:\bullet}{x},\stackrel{\:\bullet}{\lambda_{u}^{sg}}})\\
 & +\stackrel[l=1]{3}{\sum}\left[\underset{t\in T}{\sum}\underset{u\in F}{\sum}\underset{j_{t}\in V_{g}}{\sum}\mu_{l}^{i,f,t}\cdot g_{l}(\stackrel{\:\bullet}{\lambda},\stackrel{\:\bullet}{x},\stackrel{\:\bullet}{\lambda_{u}^{sg}})\right]\\
 & \leq f(\mathit{\stackrel{\:\bullet}{\lambda},\stackrel{\:\bullet}{x},\stackrel{\:\bullet}{\lambda_{u}^{sg}}})
\end{array}\label{eq:duel problem}
\end{equation}

Since (\ref{eq:duel problem}) is a linear function of $\mu$\cite{Sun2010},
the inequality relationship still holds when maximizing $\mu$. Therefore,
$d^{*}$ is the largest lower bound on $p^{*}$.

\begin{equation}
d^{*}=\underset{\left\{ \mu_{1}^{i,u,t},\mu_{2}^{i,u,t},\mu_{3}^{i,u,t}\right\} }{max}d(\mu)\leq f(\lambda,x,\lambda_{u}^{sg})=p^{*}
\end{equation}

For this reason, the lower bound of the OP can be obtained efficiently
by solving the dual problem.\qed

The expansion of (\ref{eq:Lagrange function}) is:

\begin{equation}
\begin{array}{c}
\begin{array}{c}
\begin{array}{l}
\boldsymbol{L}(\lambda,x,\lambda_{u}^{sg},\mu_{1}^{i,u,t},\mu_{2}^{i,u,t},\mu_{3}^{i,u,t})\\
=\underset{t\in T}{\sum}\underset{u\in F}{\sum}\underset{j_{t}\in V_{g}}{\sum}(-1+\mu_{1}^{i,u,t}+\mu_{2}^{i,u,t}-\mu_{3}^{i,u,t})\cdot\lambda_{u}^{sg}(i_{t},j_{t})\\
+\underset{t\in T}{\sum}\underset{u\in F}{\sum}\underset{j_{t}\in V_{g}}{\sum}\left\{ \mu_{1}^{i,u,t}-\mu_{2}^{i,u,t}+\mu_{3}^{i,u,t}\right\} \cdot\lambda_{u}(i_{t},j_{t})\\
\cdot\dfrac{(M-t)}{M}+\underset{t\in T}{\sum}\underset{u\in F}{\sum}\underset{j_{t}\in V_{g}}{\sum}\left\{ \mu_{1}^{i,u,t}\cdot\dfrac{(T_{u}^{s}-2M)}{M}\right.\\
\left.-\mu_{2}^{i,u,t}\cdot\dfrac{T_{u}^{e}}{M}+\mu_{3}^{i,u,t}\cdot\dfrac{(1+T_{u}^{e})-M}{M}\right\} 
\end{array}\end{array}\end{array}
\end{equation}

Since the complex constraints are decoupled, it can be easily observed
that for given $\mu_{1}^{i,u,t}$, $\mu_{2}^{i,u,t}$ and $\mu_{3}^{i,u,t}$,
the first term is only connected with $\lambda_{u}^{sg}(i_{t},j_{t})$,
the second term is only related to $\lambda_{u}(i_{t},j_{t})$, and
the third term is a constant. Therefore, the above formula can be
further decomposed into two subproblems:

\begin{equation}
\begin{array}{rl}
\mathbf{P1}:\underset{\left\{ \lambda,x\right\} }{min} & \underset{t\in T}{\sum}\underset{u\in F}{\sum}\underset{j_{t}\in V_{g}}{\sum}\left[\mu_{1}^{i,u,t}-\mu_{2}^{i,u,t}+\mu_{3}^{i,u,t}\right]\\
 & \cdot\lambda_{u}(i_{t},j_{t})\cdot\dfrac{(M-t)}{M}\\
s.t.: & (5)-(7),(9)-(13)\text{,}(15)-(16),\\
 & (19)-(20)
\end{array}
\end{equation}

The goal of $\mathbf{P1}$ is to maximize the number of images that
the ground station successfully receives.

\begin{equation}
\begin{array}{rl}
\mathbf{P2}:\underset{\left\{ \lambda_{u}^{sg}\right\} }{min} & \underset{t\in T}{\sum}\underset{u\in F}{\sum}\underset{j_{t}\in V_{g}}{\sum}(-1+\mu_{1}^{i,u,t}+\mu_{2}^{i,u,t}-\mu_{3}^{i,u,t})\\
 & \cdot\lambda_{u}^{sg}(i_{t},j_{t})
\end{array}
\end{equation}

The goal of $\mathbf{P2}$ is to maximize the successful transmission
ratio of time-deterministic images.

\begin{proposition} 

By solving $\mathbf{P1}$ and $\mathbf{P2}$ separately, and then
converting the obtained solutions into a feasible solution for the
OP, the resulting feasible solution provides an upper bound of the
OP.

\label{proposition: upper bound}

\end{proposition}

\textit{Proof:} Suppose $p^{*}$ is the optimal solution to the OP,
with the corresponding optimal value $f(p^{*})$. And $\hat{\boldsymbol{L}}$
is a feasible solution to the OP obtained by solving the subproblems
$\mathbf{P1}$ and $\mathbf{P2}$, with the corresponding value $f(\hat{\boldsymbol{L}})$.
Since $p^{*}$ minimizes the objective function value among all feasible
solutions, $f(\hat{\boldsymbol{L}})$ is never less than $f(p^{*})$. 

\begin{equation}
f(\hat{\boldsymbol{L}})\geq f(p^{*})
\end{equation}

Therefore, $f(\hat{\boldsymbol{L}})$ provides an upper bound for
the OP.\qed

By releasing the constraints that make the problem OP difficult, the
original problem becomes subproblems that are easy to solve. we can
obtain the solution with the Lagrange multiplier method.

\section{The 3C Resources Joint Allocation Algorithm Desgin}

The difficulty of solving OP is (\ref{eq:lamda-GS and t- -linked-1})
coupling the variables $\lambda$ and $\lambda^{sg}$. After Relaxing
it can decomposes into tractable subproblems.

For the sake of quickly obtaining a high-quality solution, we designed
the SRCC, a three-stage algorithm: The multipliers are brought into
two subproblems (using the heuristic Algorithms \ref{alg:ESA-1} and
\ref{alg:FSC}) to obtain a feasible solution to the OP. After that,
the Algorithm \ref{alg:LSA} is applied to optimize the feasible solution
and improve the upper bound. Finally, the dual problem is solved to
derive the lower bound. By updating multipliers iteratively to reduce
the distance between upper and lower bounds, the optimal solution
is progressively approached.

\subsection{The SRCC}

In this section, we desgin the SRCC to get the the approximately optimal
solution of the OP. The SRCC is executed as follows:

First, put the multipliers into the Lagrange function, and solve the
subproblems with the Algorithm \ref{alg:ESA-1} and \ref{alg:FSC},
obtaining the initial feasible solution of the original problem. Then,
the Algorithm \ref{alg:LSA} will be used to improve the quality of
the feasible solution and update the upper bound (UB). Subsequently,
put the feasible solution into DP, thereby obtaining the lower bound
(LB) by the gurobi, which is used to evaluate the quality of the feasible
solution. Afterward, calculate the dual distance: $gap=\frac{UB-LB}{UB}\times100\%$.
When the $gap\leq\varepsilon,\varepsilon>0$ or reaches the maximum
iterations $n_{max}$, the algorithm quits. Otherwise, updating the
multipliers $(\mu_{l}^{i,u,t})^{n+1}$ with the formula: \textcolor{magenta}{}

\begin{equation}
(\mu_{l}^{i,u,t})^{n+1}=max\{0,(\mu_{l}^{i,u,t})^{n}+\alpha_{l}^{n}\times\nabla_{l}^{n}\}\label{eq:update the multipliers}
\end{equation}
Where $\nabla_{l}^{n}$ is the value obtained when the solution of
the subproblem is brought into the relaxed constraint at the n-th
iteration. And $\alpha_{l}^{n}$ represents the step, which can be
obtained using the following formula:

\begin{equation}
\alpha_{l}^{n}=\dfrac{\beta\cdot(UB-LB)}{\left\Vert \nabla_{l}^{n}\right\Vert ^{2}}\label{eq:update the step}
\end{equation}
Where $\beta$ is a fixed parameter, usually $1\leq\beta_{n}<2$.
Following this, the new $(\mu_{l}^{i,u,t})^{n+1}$ is brought into
the Lagrange function for the next iteration. The algorithm terminates
when the end condition is met and returns the current approximate
optimal solution $x$, $\lambda$, and $\lambda^{sg}$. Additionally,
if the upper bound is not updated in 3 iterations, adjust $\beta=\frac{1}{2}\cdot\beta$.The
algorithm must also end if $\beta=0$, the steps are less than the
default value (such as 0.01), or the subgradients are 0.
\begin{algorithm}[tbh]
\caption{The SRCC \label{alg:TSA-RTC Algrithom}}

\small

\textbf{Input:} The MDR-TEG\textbf{ }$\mathbb{\mathbb{M}}=\left\{ V,\varepsilon,T\right\} $,
the flow set $u\in F$, the five-tuple $u=\{s,d,\delta_{u},T_{u}^{s},T_{u}^{e}\}$
.

\textbf{Output:} The $x$, $\lambda$, and $\lambda^{sg}$.

\begin{algorithmic}[1]

\STATE \textbf{Initialization: }Set the UB = +\ensuremath{\infty},
the LB = --\ensuremath{\infty}, and $n=1$. The multipliers $(\mu_{l}^{i,u,t})^{1}=1,l=1,2,3$.
The gap $\varepsilon=0$, the maximum iterations $n_{max}=100$ and
$\beta=1$.

\STATE Put the $(\mu_{l}^{i,f,t})^{n}$ into the (\ref{eq:Lagrange function}),
and solve $\mathbf{P1}$ with the Algorithm \ref{alg:ESA-1}, obtaining
the initial solution of $x$, $\lambda$ and $\lambda^{sg}$.\label{step:next interations}

\STATE Using the Algorithm \ref{alg:FSC} transform the $x$,$\lambda$
and $\lambda^{sg}$ into the feasible solution for the OP. 

\STATE The Algorithm \ref{alg:LSA} is used to enhance the quality
of feasible solutions, obtain the current upper bound Cur\_UB.

\STATE $UB=min\left\{ UB,Cur_{-}UB\right\} $.

\STATE Put $x$, $\lambda$, and $\lambda^{sg}$ into DP, obtaining
the current lower bound Cur\_LB.

\STATE $LB=max\left\{ LB,Cur_{-}LB\right\} $.

\IF{$\frac{UB-LB}{UB}\times100<\varepsilon$ || $n\geq n_{max}$
}

\STATE Go to \ref{step:end iteration}.

\ELSE

\STATE{Calculate the subgradient $\nabla_{l}^{n}$ with $x$, $\lambda$,
and$\lambda^{sg}$.}

\STATE{Update the step $\alpha_{l}^{n}$ according to equation (\ref{eq:update the step}).}

\STATE{Update the multiplier $(\mu_{l}^{i,u,t})^{n+1}$ according
to equation (\ref{eq:update the multipliers}).}

\STATE{n = n+1.}

\STATE{Go to \ref{step:next interations}.}

\ENDIF\

\RETURN the current approximate optimal solution, $x$, $\lambda$,
and $\lambda^{sg}$.\label{step:end iteration}

\end{algorithmic}
\end{algorithm}

\subsection{The Efficient Solution Algorithm (ESA) Bases 3C Resources}

\begin{algorithm}[tbh]
\caption{The ESA for Allocating 3C Resources\label{alg:ESA-1}}

\textbf{Input: }The $u=\{s,d,\delta_{u},T_{u}^{s},T_{u}^{e}\},u\in F$,
$N$ and link weights.

\textbf{Output:} The solution to the $\mathbf{P1}$: $x^{*}$ and
$\lambda^{*}$.

\begin{algorithmic}[1]

\STATE \textbf{Initialization:} $F_{suc}=F_{un}=\{\}$, $x^{*}=\lambda^{*}=Flag=0$,
and $t=1$.

\WHILE {$F\neq\phi$}\label{step: flow waits scheduling -1}

\STATE Sort all flows in ascending order by $T_{u}^{s}$, $T_{u}^{e}$
and $\delta_{u}$.\label{step:x_=00007Bu=00007D is arranged in ascending order-1}

\STATE Taking the first $u$, set $t_{u}^{'}=T_{u}^{s}$. $r_{u}$
is calculated by (\ref{eq:topo}).\textcolor{magenta}{\label{step:Taking the first flow-1}}

\IF{$T_{u}^{s}=t$}\label{step:Ts=00003Dslot t-1}

\STATE Extract links from $r_{u}$ with $w(i_{t},j_{t})\leq1/\delta_{u}$,
then use Dijkstra algorithm to plan the shortest path.\label{step:Flyd-1}

\IF{Successfully find the path for $u$}\label{step:successfully find path-1}

\IF{$Flag_{u}=0$}

\IF{Existing the earliest node $a_{t}$ in $t$ that can access
$V_{com}$: $w(a_{t},V_{com})\leq1/\delta_{u}$}

\STATE Allocate resources bases \ref{enu:save path before Vcom}.
Set $s=a$, $\delta_{u}=\delta_{u}\cdot\theta_{u}$, and $Flag_{u}=1$.
Go to \ref{step:Flyd-1}.

\ELSE

\STATE Allocate resources bases \ref{enu:save whole path in t}.\label{step: cross slot update path-1}

\IF{$l=d$}

\STATE Set $F=F-u$ and $F_{suc}=F_{suc}+u$. Go to \ref{step:Taking the first flow-1}.

\ELSE

\STATE Set $s=l$, updating $T_{u}^{s}=t+1$, then put $u$ back
in $F$. Go to \ref{step:x_=00007Bu=00007D is arranged in ascending order-1}.

\ENDIF\

\ENDIF\

\ELSE

\STATE Go to \ref{step: cross slot update path-1}.

\ENDIF\

\ELSE

\STATE set $t_{u}'=t_{u}'+1$.

\IF{$t_{u}'<=T_{u}^{e}$}

\STATE Expand the $r_{u}$ with (\ref{eq:topo}), set $d_{u}=d_{u}+N$,
Go to \ref{step:Flyd-1}.

\ELSE

\STATE $F=F-u$ and $F_{un}=F_{un}+u$. Go to \ref{step: flow waits scheduling -1}.

\ENDIF\

\ENDIF\

\ELSE

\STATE $t=t+1$, Go to \ref{step:Ts=00003Dslot t-1}.

\ENDIF\

\ENDWHILE\

\RETURN $x^{*}$, $\lambda^{*}$, and $\lambda^{sg*}$.

\end{algorithmic}
\end{algorithm}

The jointing allocation of 3C resources is a path planning problem
of flows essentially. The goal of $\mathbf{P1}$ is to maximize the
number of images successfully received by the ground station. Using
computing resources in the SGIN, as early as possible when planning
flows, can save storage and transport resources to carry more flows.
In pursuit of this, we design the ESA, as shown in the Algorithm \ref{alg:ESA-1},
which can find a better solution in a short time.

Before the planning starts, using the remaining link resources $l(i_{t},j_{t})$
between any two nodes to calculate the weight: $w(i_{t},j_{t})=\frac{1}{l(i_{t},j_{t})}$.
And excluding the flow that is unsuitable for planning. If $\mu_{1}^{i,u,t}-\mu_{2}^{i,u,t}+\mu_{3}^{i,u,t}\geq0$,
proves that programming $u$ will deviate from the optimal solution.
Then move it out of the $F$, set $F=F-u$ and $F_{fail}=F_{fail}+u$.
Meanwhile, the scheduling order must be organized reasonably. Arranging
the remaining flows of $F$ in ascending order according to $T_{u}^{s}$,
$T_{u}^{e}$ and $\delta_{u}$, so that the flow with short remaining
time can be scheduled priority. This optimizes the solutions' feasibility
in $\mathbf{P2}$. 

During the planning, calculate the shortest path for $u$ that has
sending demand in the current scheduling slot $t$, with the link
satisfying $w(i_{t},j_{t})\leq1/\delta_{u}$. If the path cannot be
found in $t$, expand the searching slot range $t_{u}'$ in MDR-TEG
by a slot (i.e., $t_{u}'=t_{u}'+1$), and continue finding the path
in the newly searching range size $r_{u}$ (\ref{eq:topo}) correspondingly. 

\begin{equation}
r_{u}=[N\cdot(t^{'}-T_{u}^{s}+1)]^{2}\label{eq:topo}
\end{equation}
wherein, $N=V_{o}+V_{s}+V_{g}$ is the sum of real nodes in the SGIN.
The $u$ will be added to the unplanned set $F_{un}$ until $t_{u}'$
exceeds $T_{u}^{e}$ still can't find a path.

After successfully finding the way of $u$, The allocating resources
directly for $u$ spanning multiple slots may compromise the scheduling
of other flows. As shown in Fig. \ref{fig:cross slots resources allocation}.
The computed cross-slot transmission path of flow 1 is $o_{1}^{1}\rightarrow s_{1}^{1}\rightarrow com\rightarrow s_{1}^{1}\rightarrow s_{1}^{2}\rightarrow s_{2}^{2}\rightarrow g_{1}^{2}$.
Pre-allocating resources across slots results in insufficient resources
for flow 2, preventing its completion within the two slots. 

The brilliance of ESA lies in its per-scheduling-slot resource allocation
approach, which retains the path within $t$ and flexibly plans the
rest. Consider the $u$, which successfully finds a cross-slot path,
as an example to demonstrate ESA's resource allocation process. This
scheme is also applicable to flows find a complete path within a single
slot. The resource allocation is categorized into two cases as follows:

\begin{figure}[t]
\begin{centering}
\includegraphics[scale=0.47]{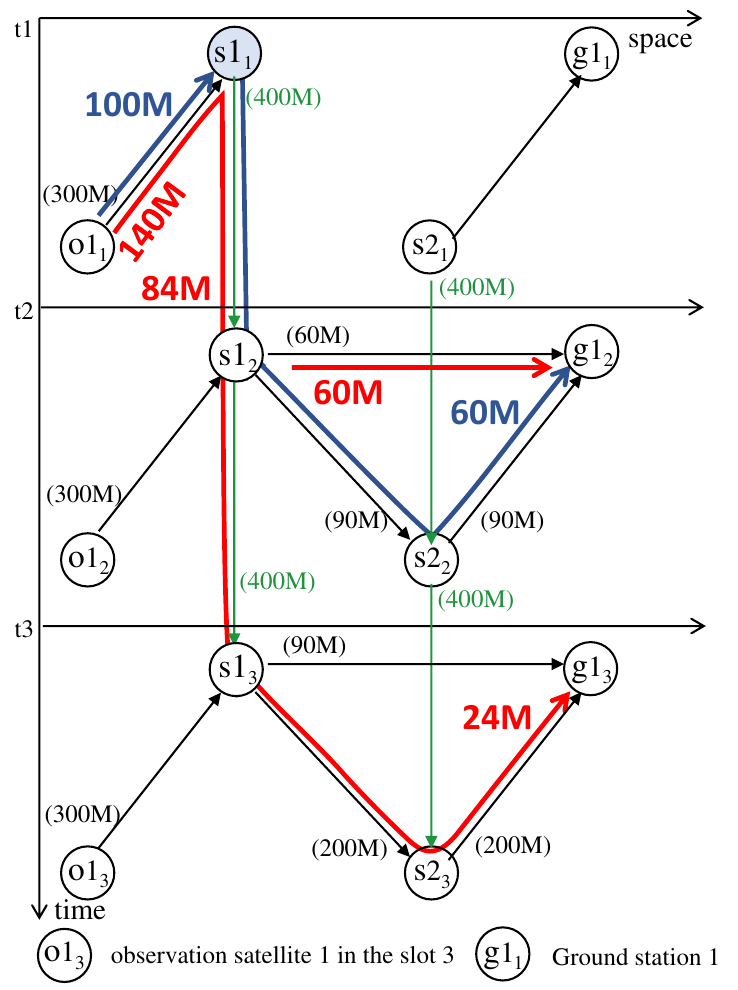}
\par\end{centering}
\centering{}\caption{The rigidity of directly allocating resources for cross slots path.\label{fig:cross slots resources allocation}}
\end{figure}

\subsubsection{Resource Allocation Strategy per Scheduling Slot Before Compression\textcolor{magenta}{\label{enu:save path before Vcom}}}

If $u$ has not been compressed, and computing resources can be allocated
to $u$ in $t$, reserve the path between $s$ and $V_{com}$. The
resources consumed on the path is: $x_{u}^{*}(i_{t},j_{t})=\delta_{u}$,
After compression, the resources consumed are $x_{u}^{*}(i_{t},j_{t})=\delta_{u}\cdot\theta_{u},(i_{t},j_{t})\in\varepsilon_{cs}$
and the compression loss is $x_{u}^{*}(i_{t},j_{t})=\delta_{u}\cdot(1-\theta_{u}),(i_{t},j_{t})\in\varepsilon_{ca}$.
Update $\lambda_{u}^{*}(i_{t},j_{t})$ and link weight $w(i_{t},j_{t})$
with (\ref{eq:x and lamuda- - linked}) and (\ref{eq:Update the link weight}).
Update source nodes and traffic $\delta_{u}=\delta_{u}\cdot\theta_{u}$,
and re-plan subsequent path. 

\begin{equation}
w(i_{t},j_{t})=\frac{1}{l(i_{t},j_{t}){\color{red}{\color{black}-}}x_{u}^{*}(i_{t},j_{t})}\label{eq:Update the link weight}
\end{equation}

\subsubsection{Resource Allocation Strategy per Scheduling Slot after compression
or lacking computing resources\textcolor{magenta}{\label{enu:save whole path in t}}}

When $u$ has been compressed or the compressed resources are insufficient,
keep the path information from $s$ to the last hop $l$ in $t$.
The resources consumed on the related path is: $x_{u}^{*}(i_{t},j_{t})=\delta_{u}$,
According to the formula (\ref{eq:x and lamuda- - linked}) and (\ref{eq:Update the link weight})
update $\lambda_{u}^{*}(i_{t},j_{t})$ and $w(i_{t},j_{t})$. Update
the source node, put it back in $F$, and plan the subsequent path
together with other flows waiting for scheduling in the $t+1$. Until
the destination is successfully planned.

\begin{figure}[t]
\begin{centering}
\includegraphics[scale=0.47]{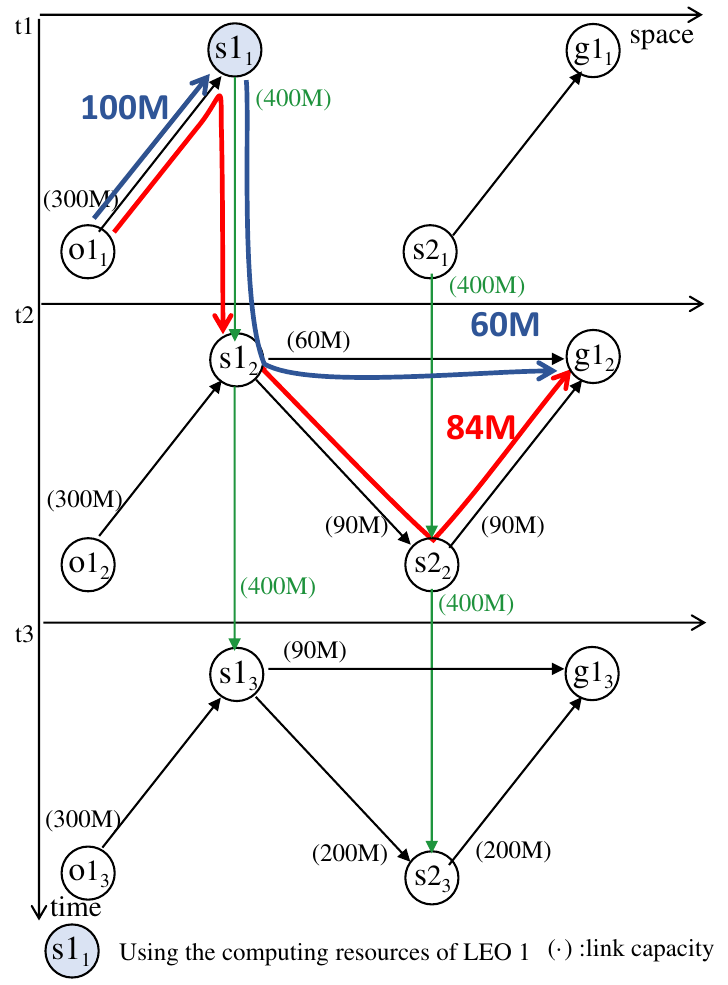}
\par\end{centering}
\centering{}\caption{Per slot Resource Allocation Strategy.\textcolor{magenta}{\label{fig:per slot resource allocation approach}}}
\end{figure}

As shown in Fig. \ref{fig:per slot resource allocation approach},
using ESA, the transmission path for flow 1 after compression is adjusted
to $s_{1}^{1}\rightarrow s_{1}^{2}\rightarrow g_{1}^{2}$. This optimization
allows flow 2 to be transmitted via the path $o_{1}^{1}\rightarrow s_{1}^{1}\rightarrow com\rightarrow s_{1}^{1}\rightarrow s_{1}^{2}\rightarrow s_{2}^{2}\rightarrow g_{1}^{2}$,
completing the scheduling within two slots. This ensures optimal resource
allocation within each $t$, thereby improving the successful transmission
ratio.

\subsection{The Feasible Solution Correction (FSC) Algorithm}

$\mathbf{P1}$ takes sending time of flows into account when planning
path, which enhance the objective function value for $\mathbf{P2}$.
The Algorithm \ref{alg:FSC} depicts the details of solving $\mathbf{P2}$
and converting the solution into a feasible solution.

Put the path planning result of $\mathbf{P1}$ into $\mathbf{P2}$,
$\lambda$ and (\ref{eq:lamda-GS and t- -linked-1}) are used to calculate
the $\lambda^{sg}$, further achieving image path planning that satisfies
delay constraints. $\lambda_{u}^{sg}(i_{t},j_{t})=0$ indicates that
despite planning a path for $u$, it still arrive later than expected
$T_{u}^{e}$. Consequently, the resources assigned to $u$ should
be released and add it to the set $F_{un}$. Considering the remaining
flows in $F_{suc}$ may still breach the relaxed constraints, eliminating
those violating flows will produce a feasible solution for the OP.
\begin{algorithm}[t]
\caption{Feasible Solution Correction (FSC) Algorithm\label{alg:FSC}}

\textbf{Input:} The $x$, $\lambda$ and $u\in F_{suc}$. The multipliers
$(\mu_{l}^{i,u,t})^{n},l=1,2,3$.

\textbf{Output:} The feasible solution to the OP. 

\begin{algorithmic}[1]

\STATE \textbf{Initialization:} $\lambda^{sg}=0$, and $sum_{u}=0$.

\FOR {each $u\in F_{suc}$ }

\STATE\ Put $\lambda_{u}(i_{t},j_{t})$ into the (\ref{eq:lamda-GS and t- -linked-1})
to calculate $\lambda_{u}^{sg}(i_{t},j_{t})$.

\IF{$\lambda_{u}^{sg}(i_{t},j_{t})=0$}

\STATE\ Release all resources that $u$ consumes on the path. $x_{u}=\lambda_{u}=0$.

\STATE\ Recover the weight of the link with $w(i_{t},j_{t})=\frac{1}{l(i_{t},j_{t})+x_{u}(i_{t},j_{t})},\forall(i_{t},j_{t})\in\varepsilon$.

\STATE\ $F_{suc}=F_{suc}-u$, and $F_{un}=F_{un}+u$.

\ENDIF\ 

\ENDFOR\

\IF{$F_{suc}\neq\phi$}

\FOR {each $u\in F_{suc}$ }

\IF{$(-1+\mu_{1}^{i,u,t}+\mu_{2}^{i,u,t}-\mu_{3}^{i,u,t})\geq0$}

\STATE set $F_{suc}=F_{suc}-u$, $F_{fail}=F_{fail}+u$.

\STATE\ Recover the weight of the link with $w(i_{t},j_{t})=\frac{1}{l(i_{t},j_{t})+x_{u}(i_{t},j_{t})},\forall(i_{t},j_{t})\in\varepsilon$.

\ENDIF\

\ENDFOR\

\ENDIF\

\STATE $sum_{u}=\underset{u\in F_{suc}}{\sum}\lambda_{u}^{sg}(i_{t},j_{t})$

\RETURN $x$, $\lambda$, $\lambda^{sg}$ and $sum_{u}$.

\end{algorithmic}
\end{algorithm}

\subsection{The Local Search Algorithm (LSA)}

To rapidly find a near-optimal solution, this section introduces the
LSA to further enhance the quality of feasible solutions within a
single SRCC iteration. Theoretically, the FSC has already removed
all unsuitable scheduling flows into the set $F_{ail}$. By adjusting
the scheduling order, flows in $F_{un}$ still have the potential
to arrive on time. Algorithm \ref{alg:LSA} reorders the remaining
flows of $F_{suc}$ and $F_{un}$, then inputs them into Algorithm
\ref{alg:ESA-1} for rescheduling. And using Algorithm \ref{alg:FSC}
to update the upper bound $Cur_{-}UB$. The algorithm iterates until
$Cur_{-}UB$ cannot be further improved. At this point, the result
obtained is considered the near-optimal solution derived from the
current SRCC.
\begin{algorithm}[t]
\caption{The LSA\label{alg:LSA}}

\textbf{Input:} $x$, $\lambda$, $\lambda^{sg}$ and $sum_{u}$.

\textbf{Output:} The $Cur_{-}UB$.

\begin{algorithmic}[1]

\STATE \textbf{Initialization:} $Cur_{-}UB=0$.

\FOR { $i=1:|F_{un}|$ }

\STATE Using the Algorithm \ref{alg:ESA-1} to replan the path of
$F_{suc}$ and $F_{un}$ respectively.

\STATE Obtain the $sum_{u}$ by the Algorithm \ref{alg:FSC}.

\IF{$sum_{u}>Cur_{-}UB$}

\STATE $Cur_{-}UB=sum_{u}$.

\ENDIF\ 

\ENDFOR\

\RETURN The $Cur_{-}UB$.

\end{algorithmic}
\end{algorithm}

\section{Simulation Results\label{sec:5 Simulation Results}}

In this section, we explore the performance of the SRCC algorithm
based on MDR-TEG with the Iridium satellite constellation system which
has 6 orbits and 11 satellites in a orbit. All the real satellite
motion trajectories are generated using STK simulation software, and
the simulation time is from 2024-08-16 04:00:00 to 2024-08-17 04:00:00.
The SGIN is also equipped with five ground stations: Kiamusze (43.83$\lyxmathsym{\textdegree}$\emph{N},
130.35$\lyxmathsym{\textdegree}$\emph{E}), Xiong'an (38.9$\lyxmathsym{\textdegree}$\emph{N},
116$\lyxmathsym{\textdegree}$\emph{E}), Korla (41.68$\lyxmathsym{\textdegree}$\emph{N},
80.06$\lyxmathsym{\textdegree}$\emph{E}), Tongchuan (34.9$\lyxmathsym{\textdegree}$\emph{N},
108.93$\lyxmathsym{\textdegree}$\emph{E}), and Hainan (19.65$\lyxmathsym{\textdegree}$\emph{N},
110.3$\lyxmathsym{\textdegree}$\emph{E}). Setting the time horizon
$\tau=300$s to derive the coordinates of the satellites within each
slot. Importing these data into the MATLAB to calculate the visible
relationship between the satellites and the ground stations, and generate
the adjacency matrix of the entire network. And selected Two satellites
randomly as observation satellites, and subsequently obtain the topology
of the SGIN.

The observation nodes are taken as the sources and the ground station
as the destinations. The image size in each source node is randomly
selected within {[}100,140{]} Mbits, $\rho=20$ Mbits/units and $\zeta_{max}=20$
units \cite{Hao2022}. Before the simulation starts, the object is
set to explore the number of images that can reach the goal within
a certain time frame and not time out. According to \cite{Liu2023},
other simulation parameters are set as shown in Table \ref{tab:SIMULATION-PARAMETERS-SETTING}.
\begin{table}[t]
\caption{SIMULATION PARAMETERS SETTING.\label{tab:SIMULATION-PARAMETERS-SETTING}}

\centering{}%
\begin{tabular}{ccc}
\hline 
\textbf{Parameter} & \textbf{Value 1} & \textbf{Value 2}\tabularnewline
\hline 
The inter-satellite link capacity & 300 Mbits & 3 Gbits\tabularnewline
The satellite-to-earth link capacity & 500 Mbits & 5 Gbits\tabularnewline
The storage link capacity & 1000 Mbits & 1 Gbits\tabularnewline
The computing link capacity & 400 Mbits & 4 Gbits\tabularnewline
Flow volume & {[}100,140{]} Mbits & 100 Mbits\tabularnewline
Delay-bound & 12 slots & 10 slots\tabularnewline
Presetted time frame & 20 slots & 20 slots\tabularnewline
\hline 
\end{tabular}
\end{table}

\subsection{Benchmark Algorithm and Model}

We compare the proposed algorithm with existing models and algorithms
to study its performance in terms of execution efficiency and resource
allocation efficiency.

\subsubsection{Exhaustive search algorithm based on Lagrange relaxation (ESALR)}

The ESALR solves the Lagrangian relaxation problem exhaustively and
finds the optimal solution among all the solutions in each iteration. 

\subsubsection{Joint allocation (JA) algorithm}

JA algorithm is based on the TEG model for joint allocation of two-dimensional
(i.e., communication and storage) resources.

\subsubsection{Computing Resources Priority Allocation Algorithm (CRPAA)}

The CRPAA allocates the computing resources on the planned cross-slot
path preferentially.

\subsection{Simulation Results Comparison}

We simulate the algorithms under two different values of network parameters.
The simulation results are shown in Fig. \ref{fig3} - Fig. \ref{fig8:D2D-delay}.

\subsubsection{The simulation results comparison under value 1}

\begin{figure}[t]
\begin{centering}
\includegraphics[scale=0.35]{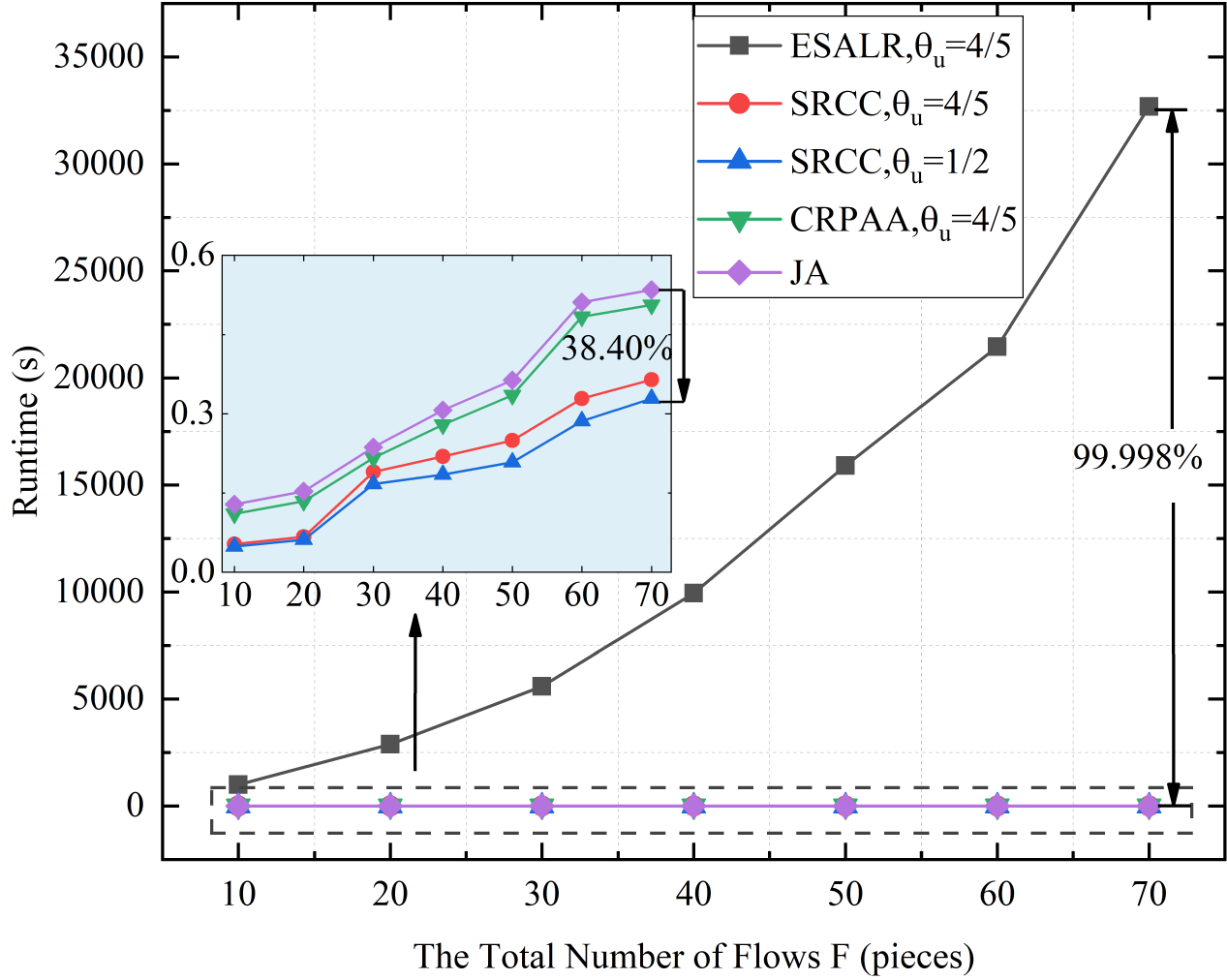}
\par\end{centering}
\centering{}\caption{The running time versus the total number of images.\label{fig3}}
\end{figure}

The Fig. \ref{fig3} illustrates the impact of the total number of
images $F$ on the algorithm's execution time. With the increase in
the total number of images, the ESALR's execution time increases almost
exponentially while the SRCC, the CRPAA, and the JA are always less
than 0.6s. Especially when the total number of images is 70 (i.e.,
$F=70$), the execution time of the ESALR algorithm reaches 32675s,
which is increased by 99.998\% compared with other algorithms. The
ESALR incurs incalculable time overhead, which is unacceptable in
a real-world scenario. We can see that the multi-dimensional resource
joint allocation (i.e., the SRCC, the CRPAA, and the JA) algorithms
can effectively reduce execution time and improve execution efficiency.
The reason is that the exhaustive algorithm will try to find every
possible solution and try all the possibilities, the others only find
a suboptimal solution for a short time. The enlarged figure shows
that the SRCC algorithm has the lowest time cost. When $F=70$ and
$\theta_{u}=1/2$, the execution time of SRCC is 38.40\% lower than
JA.

\begin{figure}[t]
\begin{centering}
\includegraphics[scale=0.35]{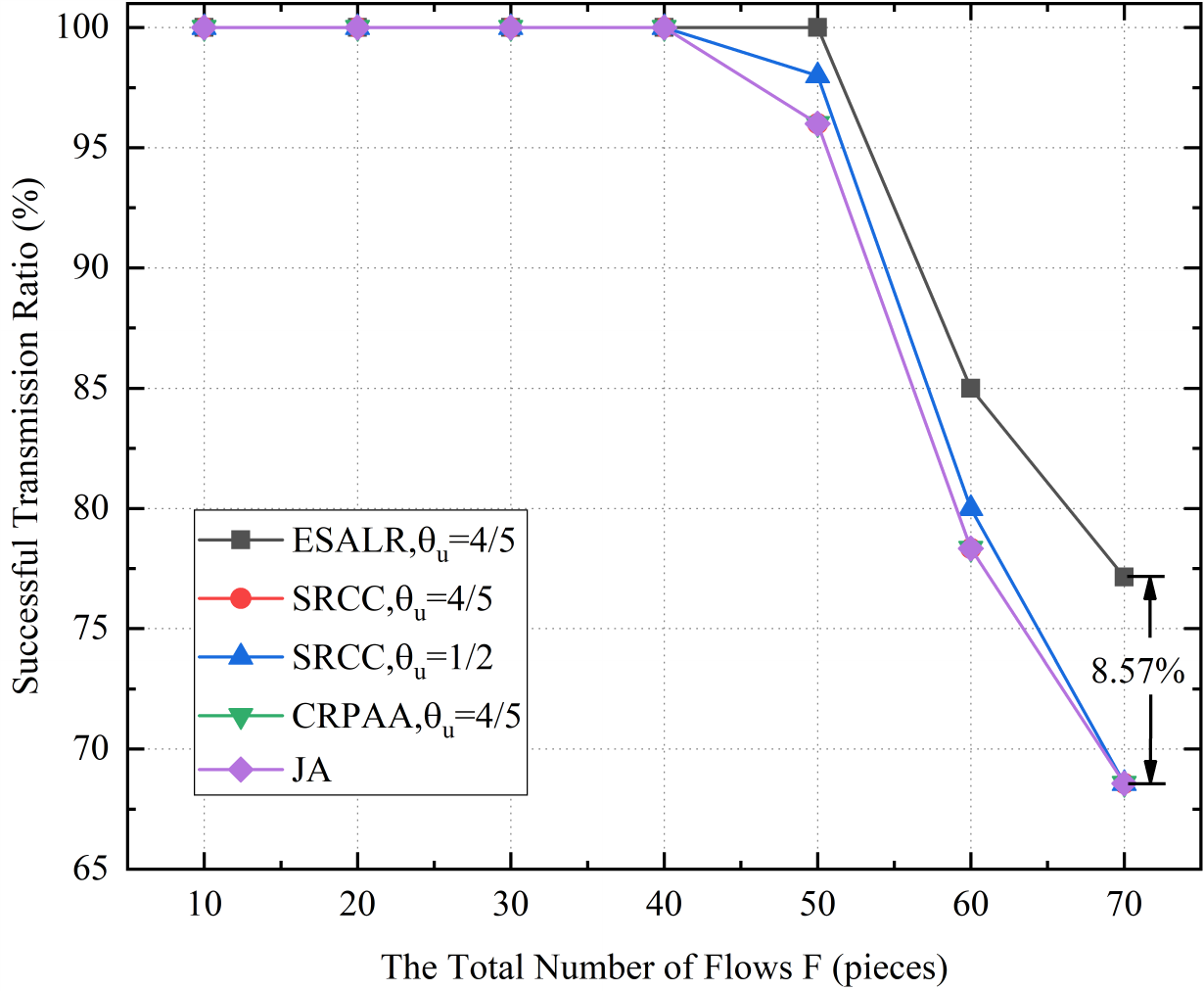}
\par\end{centering}
\centering{}\caption{Successful transmission ratio versus the total number of images.\label{fig4}}
\end{figure}

\begin{figure}[t]
\begin{centering}
\includegraphics[scale=0.35]{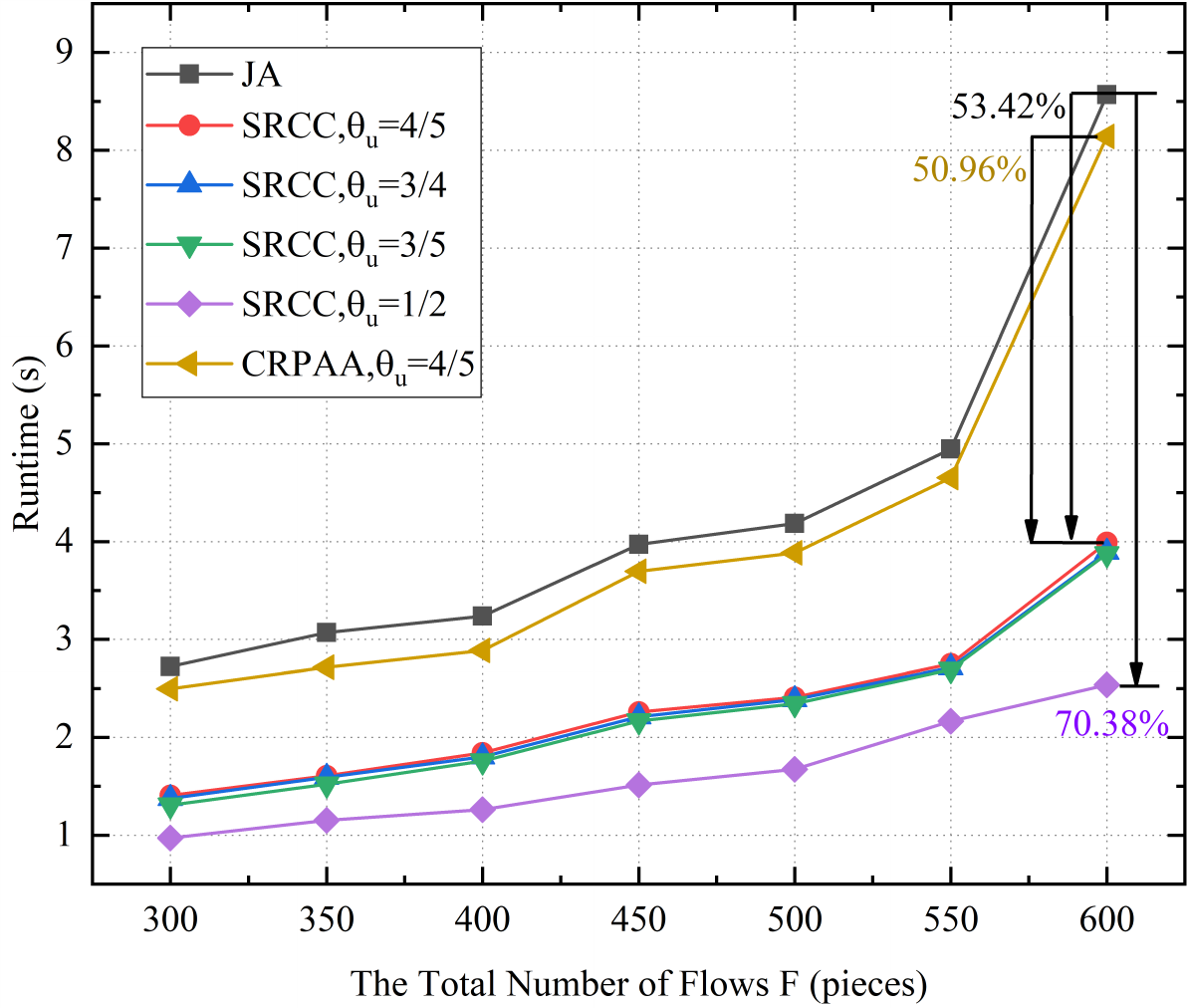}
\par\end{centering}
\centering{}\caption{The running time versus the total number of images.\label{fig5}}
\end{figure}

Fig. \ref{fig4} demonstrates how the total number of images $F$
affects the successful transmission ratio. With the increase of $F$,
the successful transmission ratio of all algorithms shows a gradual
decline. When $F\leq40$, the difference between the three algorithms
is inconspicuous, which is because there is less flow in the SGIN
and the resources are relatively sufficient. As $F$ continued to
increase, the successful transmission ratio began to show variability
due to the lack of resources. The curves of CRPAA and JA overlap and
are the lowest, while the SRCC performs better because it can re-plan
the path after calculation allowing it to better support more images.
When $F=70$, the difference between SRCC and ESALR is only 8.57\%,
and the overlap of SRCC, CRPAA, and JA is due to the $F$ is relatively
small. Therefore, we further investigate how a larger $F$ influences
algorithm performance.

\subsubsection{The simulation results comparison under value 2}

When continuing to expand the link capacity and $F$, the ESALR execution
time cannot be counted. Therefore, we further compared the performance
of SRCC with JA and different $\theta_{u}$ of CRPAA. Fig. \ref{fig5}
depicts the effect of the total number of images on runtime. We can
clearly find that as $F$ increases, the number of flows that need
to be planned in the SGIN increases, so the execution time of all
algorithms increases. Further, since JA can only allocate storage
and transmission resources, flow planning consumes the most time.
Conversely, the execution time of SRCC is clearly lower than that
of other algorithms, and it decreases further as the compression rate
$\theta_{u}$ becomes smaller. When F = 600 and $\theta_{u}=1/2$,
the SRCC's execution time significantly decreases to 70.38\% of JA's.
When $\theta_{u}=4/5$, the execution time is reduced by 50.96\% compared
to that of CRPAA under the same conditions. Because SRCC's Algorithm
\ref{alg:ESA-1} can dynamically adjust the path after compression,
available resources can be reallocated to other streams, completing
the planning in a shorter time.

The successful transmission ratio versus the total number of images
with Value 2 is plotted in Fig. \ref{fig6}. With the increase of
$F$, the images that need to be scheduled in the SGIN increase, and
the resource shortage leads to the gradual decline of the successful
transmission ratio, but the performance of SRCC is always the best.
The reasons are twofold: Firstly, the SRCC prioritizes allocating
computing resources on the planned paths, saving transmission and
storage resources to transfer more images. Secondly, the Algorithm
\ref{alg:ESA-1} flexibly adjusts the path of the image to be transmitted
in each slot, the resource utilization in the SGIN is improved and
the number of arrived images is increased. When $F=600$, the SRCC's
successful transmission ratio is 52.83\% and 45.5\% higher than JA
and CRPAA, respectively.

Fig. \ref{fig7} depicts the successful transmission ratio for different
$\theta_{u}$ values versus the total number of images. It can be
observed that the successful transmission ratio of SRCC gradually
declines as $F$ increases, but it has consistently remained above
74\%. The performance improves as the compression capacity $\theta_{u}$
increases, given a constant $F$. The reason is that the SRCC can
allocate 3C resources simultaneously. As the compression capacity
increases, more images can be processed. The reduced flow volume after
compression saves subsequent network resources, which allows the SGIN
to carry more images.\textcolor{magenta}{{} }

The Fig. \ref{fig8:D2D-delay} depicts the end-to-end delay versus
the total number of images. As $F$ increases, the end-to-end delay
of the images increases. However, the SRCC consistently maintains
the lowest delay. This is because image traffic is reduced through
the efficient allocation of computing resources for compression, allowing
the SGIN to schedule more tasks within a shorter slot. Consequently,
SRCC and CRPAA achieve better end-to-end delay than the JA algorithm.
In particular, by rearranging the path of the compressed images, SRCC
can better utilize previously underutilized resources, planning additional
images and reducing latency. As a result, SRCC outperforms CRPAA.
When $F=600$ and $\theta_{u}=1/2$, the end-to-end delay of SRCC
is 12.81\% lower than that of JA. When $\theta_{u}=4/5$, SRCC's end-to-end
delay is 10.16\% lower than CRPAA's under the same conditions.\vspace{-5em}
\begin{center}
\begin{figure}[t]
\begin{centering}
\includegraphics[scale=0.35]{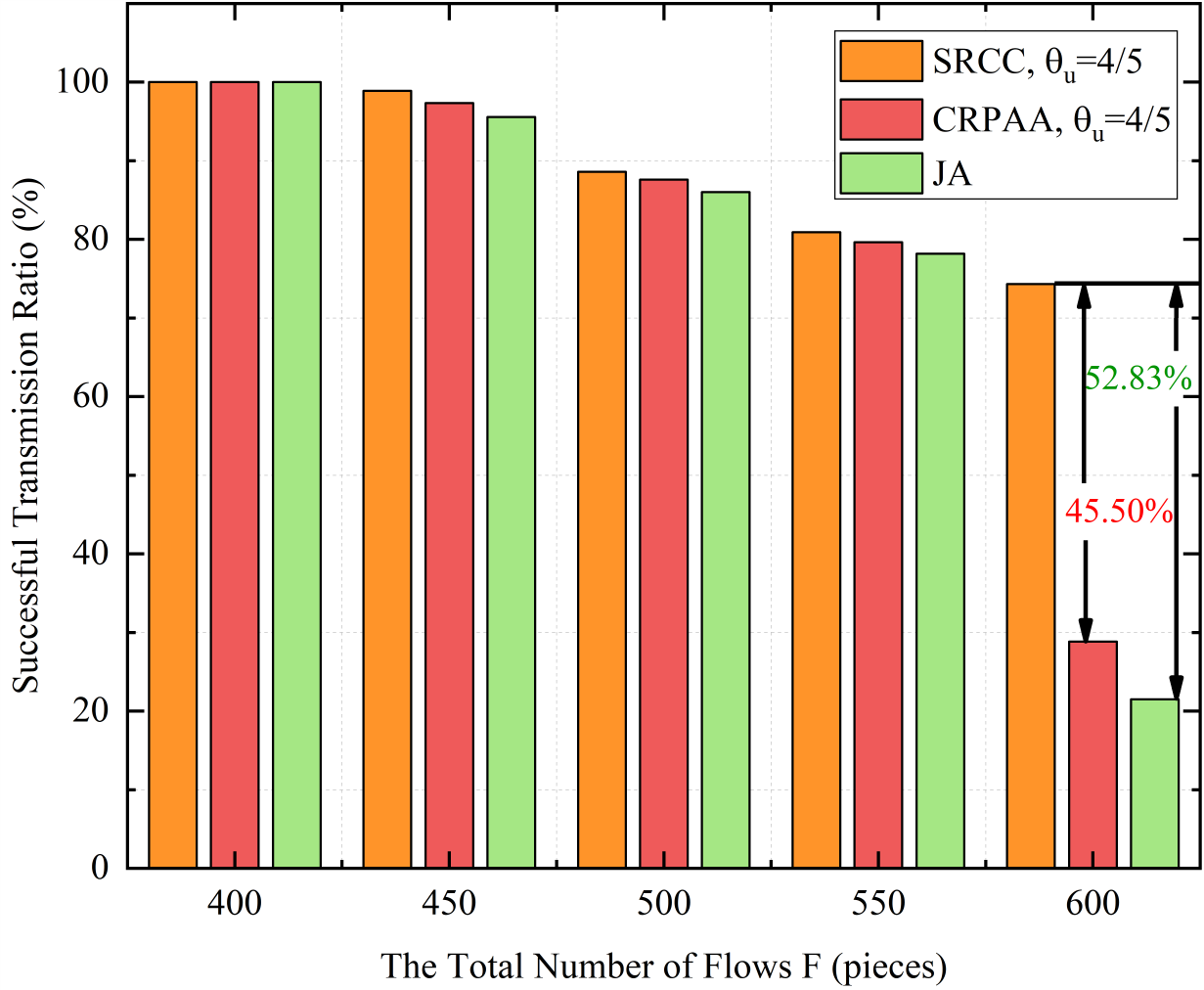}
\par\end{centering}
\centering{}\caption{Successful transmission ratio versus the total number of images.\label{fig6}}
\end{figure}
\par\end{center}

\begin{center}
\begin{figure}[t]
\begin{centering}
\includegraphics[scale=0.35]{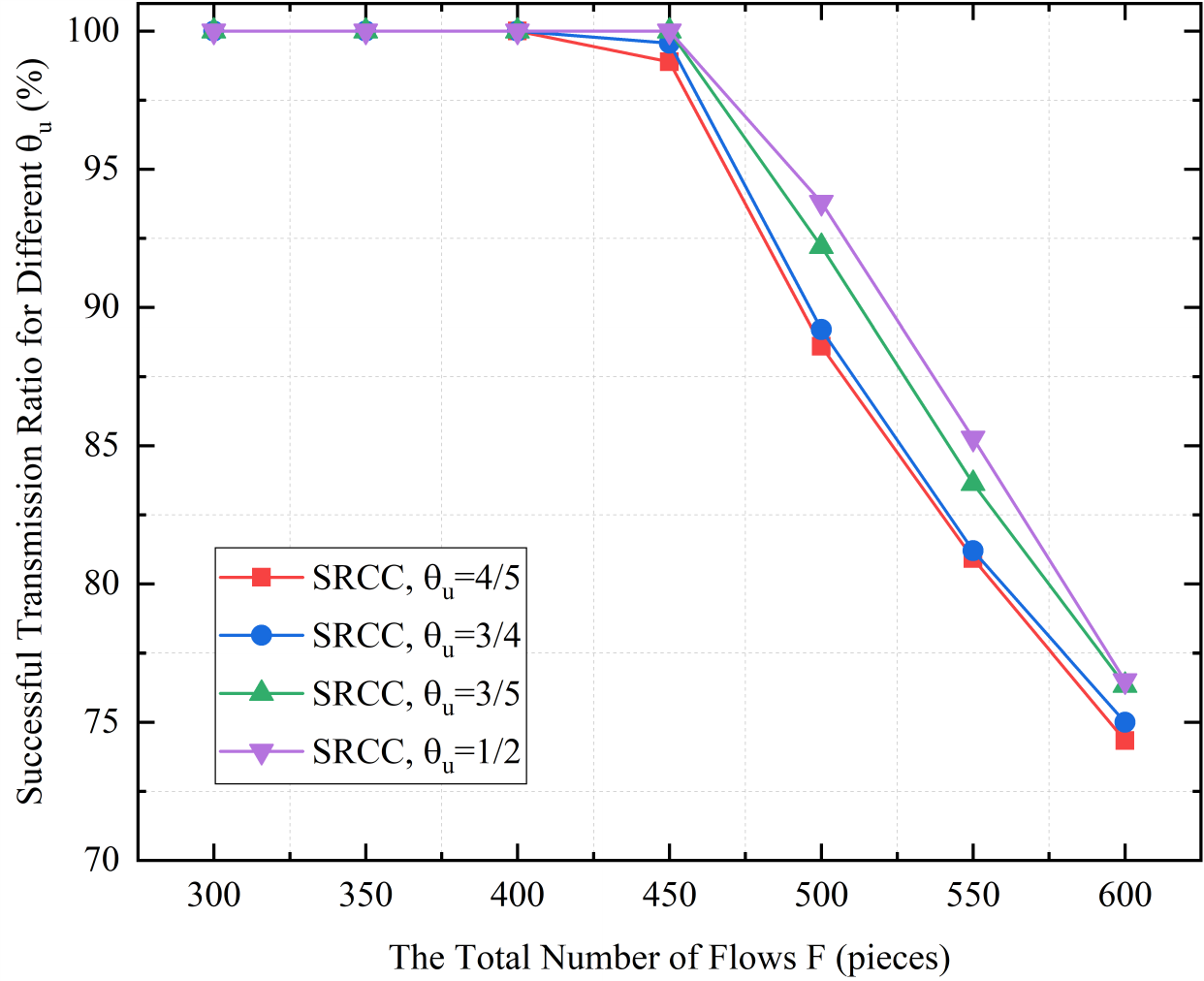}
\par\end{centering}
\centering{}\caption{Successful transmission ratio for different $\theta_{u}$ versus the
total number of images.\label{fig7}}
\end{figure}
\par\end{center}

\begin{center}
\begin{figure}[t]
\begin{centering}
\includegraphics[scale=0.35]{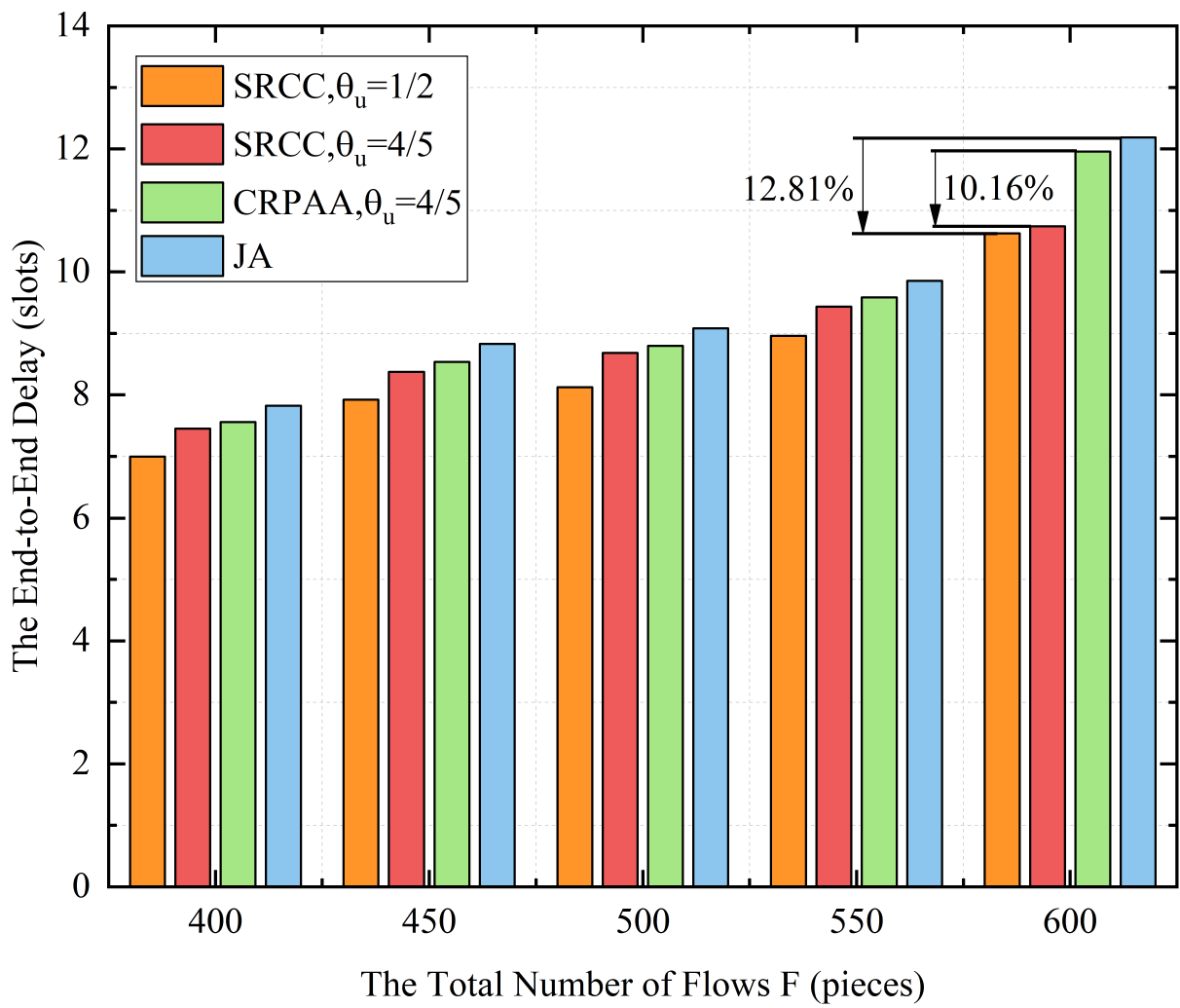}
\par\end{centering}
\centering{}\caption{The end-to-end delay versus the total number of images.\label{fig8:D2D-delay}}
\end{figure}
\par\end{center}

\section{Conclusion\label{sec:6 Conclusion}}

In this paper, we study the problem of joint allocation of storage,
computing, and transmission resources for image services. In detail,
we first design the MDR-TEG model to represent the resources in the
SGIN. Based on the MDR-TEG, our goal is to maximize the successful
transmission ratio of time-deterministic images (MSTR-TDI). Since
the MSTR-TDI is an ILP problem, which is hard. Therefore, we convert
the Lagrange relaxation problem into a dual problem and design the
SRCC algorithm to solve it. Specifically, we design the ESA algorithm
and the FSC algorithm to solve the upper bound and lower bound respectively.
The LSA algorithm is further designed to improve the quality of the
upper bound. Simulation results show that the SRCC can be executed
efficiently and brings a higher successful transmission rate. This
research can provide support for the design and optimization of the
bearing network in computing and transmission integration.

\bibliographystyle{IEEEtran}
\bibliography{Reference_long_delete}

\end{document}